\documentclass[10pt,conference,compsocconf]{IEEEtran}
\ifCLASSINFOpdf
\else
\fi
\usepackage[cmex10]{amsmath}
\usepackage{amssymb}
\usepackage{epstopdf}
\usepackage{enumerate,caption,subcaption}
\usepackage{amssymb}
\usepackage{graphicx}
\usepackage[T1]{fontenc}
\usepackage{graphicx}
\usepackage{color,verbatim}
\usepackage{multirow}
\usepackage{amsmath,amsfonts,dsfont}
\usepackage{pdfpages,cite,url}
\usepackage{fancybox}
\usepackage{balance}
\newcommand{\lambdab}{\boldsymbol{\lambda}}
\newcommand{\etab}{\boldsymbol{\eta}}
\newcommand{\pb}{\boldsymbol{p}}
\newtheorem{assumption}{Assumption}
\newtheorem{theorem}{Theorem}
\newtheorem{corollary}{Corollary}
\newtheorem{lemma}{Lemma}
\newtheorem{proposition}{Proposition}  

\DeclareMathOperator*{\argmin}{arg\,min}

\begin{document}
\title{A Game-Theoretic Study on Non-Monetary Incentives in Data Analytics Projects with Privacy Implications}

% A Game-Theoretic Proposal to Offer Users Control in Data Analytics Projects with Privacy Implications
% Collecting personal data from a repository for population estimates: a privacy protecting public good game and non-monetary incentives to modify agents' behavior

\author{\IEEEauthorblockN{Michela Chessa}
\IEEEauthorblockA{EURECOM\\
Sophia Antipolis, France\\
\url{michela.chessa@eurecom.fr}}
\and
\IEEEauthorblockN{Jens Grossklags}
\IEEEauthorblockA{The Pennsylvania State University\\
University Park, PA, USA\\
\url{jensg@ist.psu.edu}}
\and
\IEEEauthorblockN{Patrick Loiseau}
\IEEEauthorblockA{EURECOM\\
Sophia Antipolis, France\\
\url{patrick.loiseau@eurecom.fr}}
}

%\author{\ }
%\IEEEauthorblockA{School of Electrical and\\Computer Engineering\\
%Georgia Institute of Technology\\
%Atlanta, Georgia 30332--0250\\
%Email: http://www.michaelshell.org/contact.html}
%\and
%\IEEEauthorblockN{Homer Simpson}
%\IEEEauthorblockA{Twentieth Century Fox\\
%\and
%\IEEEauthorblockN{James Kirk\\ and Montgomery Scott}
%\IEEEauthorblockA{Starfleet Academy\\
%San Francisco, California 96678-2391\\



\maketitle

\begin{abstract}
%\boldmath
The amount of personal information contributed by individuals to digital repositories such as social network sites has grown substantially. The existence of this data offers unprecedented opportunities for data analytics research in various domains of societal importance including medicine and public policy. The results of these analyses can be considered a public good which benefits data contributors as well as individuals who are not making their data available. At the same time, the release of personal information carries perceived and actual privacy risks to the contributors. Our research addresses this problem area. 

%To increase the value of the derived public goods it is important to better understand the incentives of individuals for participation in the analyses projects given their privacy concerns, and the opportunities for the analyst to shape this decision-making process. Our research addresses this problem area. 

In our work, we study a game-theoretic model in which individuals take control over participation in data analytics projects in two ways: 1) individuals can contribute data at a self-chosen level of precision, and 2) individuals can decide whether they want to contribute at all (or not). From the analyst's perspective, we investigate to which degree the research analyst has flexibility to set requirements for data precision, so that individuals are still willing to contribute to the project, and the quality of the estimation improves. 

We study this tradeoff scenario for populations of homogeneous and heterogeneous individuals, and determine Nash equilibria that reflect the optimal level of participation and precision of contributions. We further prove that the analyst can substantially increase the accuracy of the analysis by imposing a lower bound on the precision of the data that users can reveal.

\end{abstract}


% \fi
\IEEEpeerreviewmaketitle

\section{Introduction}
\subsection{Background}

The seminal ``How much Information Project?'' report published in 2000 concluded that between 1 and 2 exabytes of unique information were produced worldwide per year which translated into about 250 megabytes of information for every human being \cite{hmi2000,hmi2003}. While those figures were (and are still) largely driven by commercial production of information, in recent years the amount of personal information produced by individuals has grown substantially. Now, Facebook alone absorbs about 220 petabytes of user-contributed data each year \cite{Facebook2012}. Recognizing the opportunities to economically benefit from this growth, personal data has been heralded as the ``New Oil'' of the 21st Century \cite{WEC11}. Similarly, opportunities are increasingly taken advantage of to utilize the data for research. From the individual's perspective the latter trend results in a tradeoff calculus.

%The trend to economically utilize consumer data is facilitated by the growing importance and popularity of cloud computing services and social network sites. Similarly, special purpose databases such as for the storage of direct-to-consumer genetic test results are becoming available (see, for example, \cite{Greshake14}).

On the one hand, individuals recognize that many complex challenges with societal importance, such as public health considerations, market-research or political decision-making \cite{Varian2014}, may benefit from a more rigorous analytic treatment, thanks to data analytics research and the newly-won abundance of personal information. From this perspective, many analytic results that are based on individuals' personal data can be interpreted as \textit{public goods} with societal importance. For example, advancements to better understand certain illnesses do not only potentially benefit the contributors of personal data, but are often made accessible to people in a particular domain (e.g., citizen of a country, individuals in a certain social status or demographic category, or everybody).

On the other hand, the same individuals have justified \textit{privacy concerns} about the release of their personal data. The reasons for privacy concerns can be quite diverse as outlined in Solove's privacy taxonomy \cite{Solove06}. Individuals may perceive the release and use of their data as an intrusion of their personal sphere \cite{Altman75,Warren1890}, or as a violation of their dignity \cite{Westin70}. In addition, they may fear this data can be abused for unsolicited advertisements, or social and economic discrimination (e.g., \cite{Acquisti13,Mikians13}).  

The published studies demonstrate the need to organize the collection of personal data when facing this users' tradeoff scenario, by implementing effective control and participation mechanisms. It has been shown that a majority of individuals consider it as important to be able to exercise \textit{control} over the release of their personal data \cite{Acquisti05}. For example, a number of empirical studies have provided evidence for such desires for control in the medical domain \cite{Kass03,Damschroder07,Robling04}. Moreover, even if data privacy provisions are met, many respondents would still require notice and consent over their medical data release \cite{Damschroder07,Robling04,Willison07}. Finally, several studies show a high overall concern for certain data releases. For example, a meta-review of published surveys showed that in some contexts a majority of respondents were entirely uncomfortable with health research if effective notice and consent practices were absent \cite{Westin07}. Similar findings can be shown for other problem domains.

\subsection{Problem Statement and Approach}

Our research addresses the problem area identified in the above section. 
In this paper, we propose individuals' incentives to participate in data analysis projects. These individuals face a tradeoff between having privacy cost associated with their data release, but also deriving benefits from the analysis' results. %However, the analysis requires a sufficient level of participation to determine an accurate population estimate about a particular attribute. 

We are particularly motivated by the scenario when data about individuals is already stored in a secure database for a different primary purpose (e.g., social networking or medical services). An analyst can then request the participation of individuals in a data analysis project (via a notice and consent process with negligible cost) that provides a public good. More precisely, individuals make decisions about the release of a private value given a population-relevant metric. The analyst has the objective of accurately estimating the associated population average for all individuals.

Our main focus is on understanding the incentives of individuals to participate, and of the analyst to shape this decision-making process. From each individual's perspective, control over participation takes two forms: 1) individuals can contribute data at a self-chosen level of precision, and 2) individuals can decide whether they want to contribute at all (or not). From the analyst's perspective, we investigate to which degree the research analyst has flexibility to set requirements for data precision, so that individuals are still willing to contribute to the project, and the quality of the estimation improves. 

Our work assumes that incentives for participation are non-monetary; that is, the main driver for data contributions is the interest in the derived public good. We base this assumption on the observation that direct monetary compensation for personal information has so-far received very little traction in the market for personal information, and that it meets little acceptance in consumer surveys.\footnote{While related empirical data is sparse, a survey reported that only about 25\% of the surveyed population would accept monetary compensation for personal information \cite{Acquisti05}. In contrast, offering discounts or free products/services for personal information is a common practice.}

We follow a game-theoretic approach to investigate the outlined trade-off calculus. We iteratively develop a model, where the starting point is a simplified version of the work by \cite{ioannidis2013linear}, that captures the interaction between an analyst and a set of individuals who have control over the release of information to the analyst. We conduct a rigorous analysis and derive concrete results about the precision of contributions, the quality of the population estimate, and the overall willingness to contribute to the project. 

\subsection{Contributions}

In this paper, we consider critical facets of realistic privacy decision-making, striking a good balance between model complexity and potential impact. We rigorously analyze a general model where users optimize a cost composed of an individual privacy cost and an estimation cost that captures the public good component of the analyst's estimation, both given by arbitrary functions satisfying relatively mild assumptions. In particular, we consider a general case with a continuous privacy cost function which allows users to choose a privacy level in a continuum of choices (and not simply a $0$-$1$ choice). We first analyze the homogeneous agents case, and then we extend our results to the case of heterogeneous agents, providing in detail the actions the analyst should take in order to improve the estimation. Evidence that privacy concerns are heterogeneous is a particularly central cornerstone of the privacy literature \cite{Spiekermann01}, and such an extension is fundamental for the applicability of the model. 

For both the homogeneous and the heterogeneous case, we determine Nash equilibria indicating the number of contributors and the optimal contribution levels by the individuals. We further prove that the analyst can increase the population estimate's accuracy simply by imposing a lower bound on the precision of the data that users can reveal (i.e., by restricting the level of precision of data contributions). While, for a fixed population of users providing data, increasing the precision of each data point clearly improves the population estimate's precision, the surprising and important aspect of our result lies in that the scheme remains incentive compatible, i.e., users are still willing to provide data with a higher precision rather than dropping out. 
We also show how to tune the minimum precision level the analyst should set in order to optimize the population estimate's accuracy. In our numerical simulations, we find a maximum improvement of the population estimate's accuracy in the order of $20-40\%$.

We further provide extensions of our modeling framework. First, we discuss a two-stage game in which the analyst may first recruit participants that commit to provide private data with a minimum precision; and only in a second stage, these agents would be asked to disclose their information. This captures scenarios in which agents are recruited for specific studies. Second, we also address the issue of costly acquisition of agents and their data for analysis purposes. While the no-cost-per-agent assumption we make throughout the remainder of the paper is a standard approach in most of the literature on public goods, we believe that certain practical scenarios require the appreciation of cost considerations, and this extension further completes our framework.

Our results provide a widely applicable method to increase the provision of a public good above voluntary contributions, simply by restricting the agents' strategy spaces. This method is attractive by its simplicity compared for instance to other schemes that involve monetary transfers; and could find utilization in other public good contexts. 

Understanding the trade-off between privacy, the quality of data analysis results, and willingness-to-participate in such projects is of current and growing importance. Analysts should not rely on overly broad or ineffective (take-it-or-leave-it) notice and consent procedures that do not accurately reflect individuals' preferences. In many privacy-sensitive scenarios such as involving medical data it is particularly unethical to deprive individuals of their opportunities to make decisions about their data, and whether they want to be involved in certain analysis projects. However, better insights about the involved incentive structures are needed to guide public policy and advancements of privacy-aware data analysis.

%A preliminary version of this work is the topic of a short paper \cite{Chessa15a}, in which results for a simplified model are provided. Assumptions there are such as monomial privacy cost and linear estimation cost, uniquely for the homogeneous case. Even if very restrictive, this model provided some fundamental results about the possibility for the analyst of improving the estimation, without providing what concrete action should be taken. With Section~\ref{specialcase}, we also further improve the analysis of this simplified scenario (as stated by \cite{Chessa15a}) as a special case; however, we primarily use this subsection to illustrate and plot concrete examples for our results.

Preliminary versions of some of the results presented in this paper appeared in our short paper \cite{Chessa15a}, in the context of a simplified model with monomial privacy cost, linear estimation cost and homogeneous agents. Here, we provide results for the general framework introduced above that relaxes such assumptions, we provide detailed results of practical importance on how the analyst should optimally selected the minimum precision level, and we provide several further extensions. In Section~\ref{specialcase}, we also provide more detailed results in the simplified setting of \cite{Chessa15a}, to qualitatively illustrate the results of the present paper.

\subsection{Roadmap}

Our paper is structured as follows. In Section~\ref{sec:related}, we review related work. We develop and describe our model in Section~\ref{mod}. We conduct our analysis in detail in Section~\ref{hom} on a canonical case of homogeneous agents. We extend the results to heterogeneous agents in Section~\ref{het}. We discuss extensions to our model in Section~\ref{extensions}, and conclude in Section~\ref{conclusions}. All proofs are relegated to the Appendix.
%~\ref{appendix}.

%
%for the purpose or population estimates, while individuals have the opportunity to vary the precision of their contribution based on their privacy preferences.
%
%\textbf{WORKING ON THIS:} In our work, we contribute to the question of how to account for individuals' privacy concerns and their desire to control release of personal and private data when an analyst would like to utilize their data for the computation of an estimate of a particular attribute in the observed population. More precisely, we are investigating the incentives to participate in surveys for the purpose or population estimates, while individuals have the opportunity to vary the precision of their contribution based on their privacy preferences.
%
%iteratively build a model to answer these questions... theoretical approach

%\textbf{ROADMAP:} Our paper is structured as follows: ....

%- how good the estimate is, whether it can be improved...?
%
%The question is important since in many contexts (such as medical and clinical information) consumer consent for secondary data analysis is likely required. However, it is unclear 
%
%
%
%
%
%how to harvest it? Not everything can be done without consumer involvement...
%P3P, other languages that set rules for data use
%
%also question on whether consent should be given?
%
%small examples: location privacy (which granularity of data do you want to reveal)?
%
%specific scenario that users are interested in a public good (the estimation output) but also are concerned about the contribution of their data
%
%that is common and misaligned interests (see game theory book)

\section{Related Work}\label{sec:related}
Our model draws on different lines of research including work on privacy in the context of data analytics, and game theoretic and public goods models. We also briefly review technical and cryptographic approaches, and behavioral research on control and data sharing.

Research on the optimal design of experiments assumes that already the stage of data collection can be influenced by the analyst in order to improve the learning of a linear model \cite{pukelsheim2006optimal,atkinson2007optimum}. In this paper, we allow the analyst to require data contributions at a certain level of precision to improve the computation of a population estimate, which is a related concept. Optimal design of experiments has been studied from the perspective of incentives \cite{horel2013budget}, or with the scope of obtaining an unbiased estimator \cite{RothSchoenebeck}. We propose to improve the design of experiments focusing on the privacy concerns of the agents.
 
Privacy-preserving techniques in the context of data analytics have a long history. Some recent papers propose new approaches, which allow users to protect their privacy selling aggregates of their data \cite{Riederer,Bilogrevic}. The more classical framework of $\epsilon$-differential privacy \cite{Dwork06,kifer2012private}, assumes that data are perturbed after an analysis has been conducted on unmodified inputs. That is, the analyst is considered trustworthy. In this framework, researchers have also studied the role of incentives \cite{ghosh-roth:privacy-auction,approximatemechanismdesign,roth-liggett,Chen13}. Our work differs, as we assume agents to be releasing their data independently, and an untrusted data analyst which motivates perturbations of data before submission. The idea of affecting the level of precision of released personal data, adding noise in advance of data analysis has been studied in the context of privacy-preserving data-mining (see, e.g., \cite{vaidya2005privacy,domingo2008survey}) and specific application scenarios such as building decision trees \cite{agrawal2000privacy}, clustering \cite{oliveira2003privacy}, and association rule mining \cite{atallah1999disclosure}. More recently, bounds have been derived on generic information-theoretic quantities and statistical estimation rates under a \emph{local privacy} model which preserves the privacy of agents even from the learner (similarly to adding noise before revealing data) \cite{Duchi13a}. 

Recent work has also studied the combinatorial optimization problem when an analyst may buy unbiased samples of data from different providers with given but potentially heterogeneous variance-price combinations \cite{Cummings15}. 
%Essentially, the idea is to compensate a provider for data collection efforts that impact the variance of the data collected, e.g., by financing a bigger sample the variance can be impacted favorably. 
In another recent working paper, analysts can access unbiased samples of private data by compensating data subjects for their data release according to their preferences \cite{Aperjis14}. Those studies are complementary to our work in which data subjects individually decide in a game-theoretic framework on the degree of data accuracy given a trade-off between their privacy and the determination of a socially valuable population estimate.

%Generic techniques to derive minimax bounds of population estimates have recently been proposed in \cite{Duchi13a}. Research has also focused on the back-and-forth between perturbing data for public release \cite{traub1984statistical,duncan2000optimal}, and reconstructing original distributions of data after such publication \cite{agrawal2000privacy}.
From a mechanism design perspective, scenarios have been studied where survey subjects are assumed to potentially misreport their private values \cite{dekel2010incentive,perote2004}, however, these behaviors are not studied in the context of a non-cooperative scenario. 
%A strategic approach is followed in \cite{ioannidis2013linear}, where an analyst performs a linear regression based on users' perturbed data (our starting point is a simplified version of this model). We continue this line of research by studying the benefits of restricting potential perturbation on the population estimate accuracy, and the incentives for participation in a game-theoretic framework.
A mechanism design perspective is taken in \cite{Cai14a} where the authors introduce monetary payments to create incentives for agents to give high quality data. Here, we do not consider monetary payments. A strategic approach is followed in \cite{ioannidis2013linear}, where an analyst performs a linear regression based on users' perturbed data. The authors in \cite{ioannidis2013linear} treat the estimation accuracy as a public good and study the equilibrium accuracy achieved without introducing monetary payments and the resulting price of anarchy. Our starting point is a simplified version of the model in \cite{ioannidis2013linear}. We continue this line of research by studying the benefits of restricting potential perturbation on the population estimate accuracy, and the incentives for participation in a game-theoretic framework.

Our research is also relevant to the context of the provisioning of public goods \cite{Morgan00a}. Our results show a new way of increasing the public good provision by restricting the agents' possible actions, as opposed to using monetary incentives. In addition, studies on interdependent privacy which capture the idea that data sharing by one agent impacts the privacy of other connected agents is complementary to our work \cite{Biczok13,pu2014economic}. We model the scenario when sharing creates privacy risks for individuals, but positive benefits for all agents.

The aforementioned theoretical works are complemented by technical approaches (which do not utilize insights from game-theory) such as secure hardware-based private information retrieval which can be applied, for example, in the context of online behavioral advertisement \cite{Backes12}; see also other approaches for privacy-preserving online targeted advertisements \cite{Guha11a}. Similarly, multi-party secure computation has been used to facilitate the fitting of logistic regression when data are held by separate parties \cite{Nardi12}, and homomorphic encryption has been applied to the scenario of linear regression \cite{Hall12}. Secure-computation notions of privacy have also been used in combination with game theory for privacy-preserving mechanism design \cite{Naor99a, Izmalkov05a}. 

To facilitate the privacy negotiation process between a data subject and an analyst, different technical protocols have been proposed. Several works are connected to the Platform for Privacy Preferences Project (P3P) which offers a protocol allowing data collectors (e.g., websites) to declare their intended use of information they collect about data subjects \cite{Cranor02}, and also provides agent tools for the user to manage those data requests \cite{Berthold01,Cranor02b}. More recent work, for example, addresses specific problem areas such as personalization \cite{Wang07}. Those mechanisms allow for user-specified policies regarding participation, but also minimum requirements for (not necessarily truthful) data sharing as specified by the analyst.

Research on user preferences and behaviors with respect to privacy has produced several results relevant to the context of our work. A survey study has shown that over 90\% of the respondents agreed with the definition of privacy as control of personal information \cite{Acquisti05} which presumably would include an interest to decide over the participation in data analysis projects. In hypothetical scenarios, individuals typically report high attitudinal valuations for their private data \cite{Acquisti12}. However, in experiments with actual private data transfers researchers have observed low thresholds for the release of such data in exchange for free services/goods or discounts \cite{Acquisti13b,Grossklags07,Spiekermann01}. A root cause for this privacy dichotomy is the complexity of understanding personal information exchanges and their consequences \cite{Acquisti05}.

The intricacies of human decision making have also been studied specifically focusing on the notion of control over information exchanges. Laboratory and online experiments have shown that control options have to be added with care to practically relevant scenarios \cite{Acquisti13c,Brandimarte12,Wang11}. For example, such options can elevate individuals' propensity to engage in riskier disclosures because their mere presence can contribute to a lowering of concerns over privacy \cite{Acquisti13c}. Another experimental study found that allowing individuals to customize personal data exchanges does not increase the number of transactions even though individuals were able to exclude unwanted aspects of those transactions \cite{Wang13}. Overall the understanding of the involved attitudes and behaviors is still work in progress. In our paper, we propose a process that is relatively straightforward to implement and to understand from a user perspective. However, approaches that fully accommodate the stated behavioral hurdles remain the subject of future work for behavioral as well as theoretical scientists.

\section{The Model}\label{mod}
In this section, we present our model in detail. We describe the strategic interaction between the individuals (which we also refer to as agents), whose information is contained in a data repository, and how the analyst, wishing to observe the data and to perform a statistical analysis, may modify the estimation by varying selected parameters. The linear model approach we take here builds on the work of \cite{ioannidis2013linear}.

\subsection{The Data Repository of Personal Data}

Let $N=\{1,\ldots ,n\}$ denote the set of agents, whose personal data are contained in the data repository. In particular, we suppose that each agent $i\in N$ is associated with a \textit{private variable} $y_i \in \mathbb R$, which contains sensitive information. Throughout our analysis, we suppose that there exists $y_M \in \mathbb R$, s.t., the private variables are of the form
\begin{align}
y_i=y_M +\epsilon_i,\	\	\forall i\in N,
\end{align}
where $\epsilon_i$ are i.i.d., zero-mean random variables with finite variance $\sigma^2 < \infty$, which capture the inherent noise. We stress that we make no further assumptions on the noise; in particular, we do not assume it is Gaussian. As a result, our model applies to a wide range of statistical inference problems, even cases where the distribution of variables is not known.

Paramter $y_M$ represents the \textit{mean} of the private variables $y_i$, and its knowledge is valuable to the analyst, for example as it allows him to predict the private variable of any agent whose data cannot be known (because it is not contained in the repository at that given moment, kept private by its owner, is not accessible due to limited computing resources, etc.). The analyst wishes to observe the available private variables $y_i$ and to compute their average as an estimation of $y_M$. In our model, we suppose that the analyst does not know the mean $y_M$, that he wishes to estimate, but he knows the variance $\sigma^2$. We argue that observing the variability of an attribute in a population is easier than estimating the mean, both for the analyst and for the population (in \cite{Huberman05}, for example, the authors show how individuals value their age and weight information according to the relative variability). 

\subsection{The Precision and the Analyst's Estimation}

We suppose that the analyst cannot directly access the private variables, rather she needs to ask the agents for their consent to be able to retrieve the information. As such, the agents have full control over their own private variables, and they have the choice to authorize or to deny the analyst's request. In particular, if wishing to contribute, but concerned about privacy, an agent can authorize the access to a perturbed value of the private variable. The \textit{perturbed variable} has the form $\tilde{y}_i=y_i+z_i$, where $z_i$ is a zero-mean random variable with variance $\sigma_i^2$. We assume that the $\{z_i\}_{i\in N}$ are independent and are also independent of the inherent noise variables $\{\epsilon_i\}_{i\in N}$. In practice, the agent chooses a given \textit{precision} $\lambda_i$ which corresponds to the inverse of the aggregate variance (inherent noise, plus artificially added noise) of the perturbed variable $\tilde{y}_i$, i.e.,
\begin{align*}
\lambda_i=1/(\sigma^2+\sigma_i^2)\in [0,1/\sigma^2],\	\	\forall i\in N.
\end{align*}
In the choice of the precision level, we have the following two extreme cases:
\begin{itemize}
\item[\textit{(i)}] when $\lambda_i= 0$, agent $i$ has very high privacy concerns. This corresponds to adding noise of infinite variance or, equivalently, this represents the fact that agent $i$ denies the access to her data;
\item[\textit{(ii)}] when $\lambda_i= 1/\sigma^2$, agent $i$ has very low privacy concerns. This corresponds to authorizing the access to the real private variable $y_i$, without adding any additional noise to the data.
\end{itemize}
The strategy set $[0,1/\sigma^2]$ contains all the possible choices for agent $i$: \textit{denying}, \textit{authorizing}, or any \textit{intermediate level} of precision (which captures a wide range of privacy concerns as documented in behavioral studies \cite{Spiekermann01}). We denote by $\lambdab=[\lambda_i]_{i\in N}$ the vector of the precisions.

Once each agent $i\in N$ has made her choice about the level of precision $\lambda_i$ and, consequently, the perturbed variable $\tilde{y}_i$ has been computed, the analyst has access to both the set of precisions and the set of perturbed variables. Then, the analyst estimates the mean as
\begin{align}\label{mean}
\hat{y}_M(\lambdab)=\frac{\sum_{i\in N}\lambda_i \tilde{y}_i}{\sum_{i\in N}\lambda_i},
\end{align}
where perturbed variables with higher precision (i.e., smaller variance) receive a larger weight. This estimator is the standard \textit{generalized least squares estimator}. It minimizes a weighted square error in which the $i$-th term is weighted by the precision of the perturbed variable $\tilde{y}_i$. This estimator is unbiased, i.e., $\mathbb E[\hat{y}_M]=y_M$, and has variance
\begin{align}\label{MeanVar}
\sigma_M^2(\lambdab) = \mathbb E[(\hat{y}_M(\lambdab)-y_M)^2] = \frac{1}{\sum_{i\in N}\lambda_i}\in [\sigma^2/n,+\infty].
\end{align}
In our model, the analyst aims at estimating the mean $y_M$, e.g., to be able to predict some additional private variables. Then, it is reasonable to assume that the analyst would use this estimator, as it is ``good'' for several reasons. In particular, it coincides with the \textit{maximum-likelihood estimator} for Gaussian noise and, most importantly, it has minimal variance amongst the linear unbiased estimators for arbitrary noise distributions. 

In the estimation, we have the following two extreme cases:
\begin{itemize}
\item[\textit{(i)}] when $\lambda_i=0$ for each $i\in N$, the variance \eqref{MeanVar} is infinite. This corresponds to the situation in which each agent denies the access to her data, and then the analyst cannot estimate $y_M$;
\item[\textit{(ii)}] when $\lambda_i=1/\sigma^2$ for each $i\in N$, the analyst estimates $y_M$ with variance $\sigma^2/n$, resulting only from the inherent noise. This corresponds to the situation in which each agent is authorizing the access to her data with maximum precision, i.e., no agent is perturbing her private variable. 
\end{itemize}
For any level of precision in $[0,1/\sigma^2]^n$, the estimated variance will be in $[\sigma^2/n, +\infty]$. The set of precision vectors for which the estimator has a finite variance is $[0,1/\sigma^2]^n\setminus \{(0,\ldots ,0)\}$. 

\subsection{The Estimation Game $\Gamma$}\label{estgame}

We next describe the interaction between the agents that results in their choices of precisions. We assume that each agent $i\in N$ wishes to minimize a cost function $J_i:[0,1/\sigma^2]^n \rightarrow \bar{\mathbb R}_+$, s.t., for each $\lambdab \in [0,1/\sigma^2]^n$,
\begin{align}\label{costfunction1stage}
J_i(\lambda,\lambdab_{-i})= c_i(\lambda_i) + f(\lambdab),
\end{align}
where we use the standard notation $\lambdab_{-i}$ to denote the collection of actions of all agents but $i$. The cost function $J_i$ of agent $i\in N$ comprises two non-negative components. The first component $c_i:[0,1/\sigma^2] \rightarrow \mathbb R_+$ represents the privacy attitude of agent $i$, and we refer to it as the \textit{privacy cost}: it is the (perceived or actual) cost that the individual incurs on account of the privacy violation sustained by revealing the private variable perturbed with a given precision. The second component $f: [0,1/\sigma^2]^n \rightarrow \bar{\mathbb R}_+$ is the \textit{estimation cost}, and we assume that it takes the form $f(\lambdab)=F(\sigma_M^2(\lambdab))$ where $F: [\sigma^2/n, +\infty)\to \mathbb R_+$ if the variance is finite, and $+\infty$ otherwise. It represents how well the analyst can estimate the mean $y_M$ and it captures the idea that it is not only in the interest of the analyst, but also of the agents, that the analyst can determine an accurate estimate of the population average $y_M$. 

In our model, the accuracy of the estimate can be understood as a public good, to which each user contributes with her choice of precision $\lambda_i$, at a given privacy cost. From this perspective, the assumption that the estimation cost is the same for all agents mirrors the usual standard assumption in the public good literature. Throughout our analysis, we make two additional assumptions:
% although users have different costs of contribution, they have the same valuation for the public good (see, e.g., \cite{Morgan00a} and references therein). 

\begin{assumption} \label{privacyassumption} 
The privacy costs $c_i:[0,1/\sigma^2]\to \mathbb R_+$, $i\in N$, are twice continuously differentiable, non-negative, non-decreasing, strictly convex and s.t. $c_i(0)=c_i'(0)=0$.
\end{assumption}

\begin{assumption} \label{estimationassumption} 
%Function $F: [0,1/\sigma^2]^n\setminus \{(0,\ldots ,0)\} \to \mathbb R_+$ is twice continuously differentiable, non-negative, non-decreasing and strictly convex.
Function $F: [\sigma^2/n, +\infty)\to \mathbb R_+$ is twice continuously differentiable, non-negative, non-decreasing and strictly convex.
\end{assumption}

To describe the strategic interaction between the agents, we define the \textit{estimation game} $\Gamma=\left\langle N,[0,1/\sigma^2]^n, (J_i)_{i\in N} \right\rangle$ with set of agents $N$, strategy space $[0,1/\sigma^2]$ for each agent $i\in N$ and cost function $J_i$ given by \eqref{costfunction1stage}.

\subsection{The Modified Estimation Game $\Gamma(S,\eta)$}\label{submodel}

As we shall see (Section~\ref{homogeneous.Gamma}), game $\Gamma$ has a unique Nash equilibrium for which the variance of the estimation is larger than the optimal one ($\sigma^2 / n$) due to the excess noise added by agents to protect their privacy. We further investigate the situation in which the analyst can modify the game and try to mitigate the effect of agents' privacy concerns in order to reduce the estimation cost (i.e., to improve the accuracy of the estimation obtained). Specifically, the analyst can implement the following two variations of the model. First, she can choose a \emph{minimum precision level} $\eta\in [0,1/\sigma^2]$, which is equivalent to fixing a maximum variance for the noise that agents can add to perturb their data. As it is not practically possible to force agents to authorize the access to their data with a given precision, we still assume that the agents can choose to deny the authorization, which is equivalent to selecting a precision level equal to zero. Second, the analyst can request the access to the personal data to only a subset $S\subseteq N$ of agents, with $s=|S|$ (for example, excluding those agents who are the most concerned about privacy).
% We see how this idea of imposing a minimum precision level allows the analyst to improve the estimation. 

In the modified game, the agents are informed of the subset of individuals who are asked to reveal their personal data, and of the minimum precision level $\eta$. They choose their precision $\lambda_i$ in the range imposed by the analyst $[\eta,1/\sigma^2]$ or decide to deny the access, i.e., select their precision equal to $0$. To analyze the strategic interaction between the agents in this variation, we define the game $\Gamma(S,\eta)=\left\langle S,\big[\{0\}\cup [\eta,1/\sigma^2]\big]^s, (J_i)_{i\in S} \right\rangle$ (where the cost function $J_i$ is still given by \eqref{costfunction1stage}), which is identical to $\Gamma$, except for the restricted set of agents and the restricted strategy space.

Observe that the original game $\Gamma$ is a special case of this modified game $\Gamma(S,\eta)$, when $S=N$ and $\eta=0$. We analyze the games $\Gamma$ and $\Gamma(S,\eta)$ as \textit{complete information games} between the agents, i.e., we assume that the set of agents, the action sets (in particular, when present, the value of the parameter $\eta$) and the costs are known by all the agents.

\section{The Homogeneous Agent Case}\label{hom}
In this section, we detail the analysis in the symmetric case where all the agents have identical privacy concerns, i.e., we assume that the privacy cost functions of all agents are the same: $c_i(\cdot)=c(\cdot)$ for each $i\in N$. This special case highlights the key aspects of our approach and provides some interesting preliminary results that yield intuitive interpretations. We will generalize our results to the heterogeneous case in Section \ref{het}. 

\subsection{The Estimation Game in the Homogeneous Case}\label{homogeneous.Gamma}

We first analyze the estimation game $\Gamma$, in which all the agents in $N$ are playing and the analyst allows them to choose any precision level between $0$ and $1/\sigma^2$. A \textit{Nash equilibrium} (in pure strategy) of this game is a strategy profile $\lambdab^*\in [0,1/\sigma^2]^n$ satisfying
\begin{align}\label{NEonestage}
\lambda_i^* \in \argmin_{\lambda_i\in [0,1/\sigma^2]} J_i(\lambda_i,\lambdab_{-i}^*),\	\	\forall i\in N.
\end{align}

The game $\Gamma$ with strategy space $[0, 1/\sigma^2]$ is a special case of the game in \cite{ioannidis2013linear}, where the existence of a unique Nash equilibrium is established. However, our specific assumptions allow us to characterize the equilibrium in more detail: 
\begin{theorem}\label{theorem1stage}
The game $\Gamma$ has a unique Nash equilibrium $\lambdab^*$ s.t. $\lambda_i^*=\lambda^*>0$ for each $i\in N$.
\end{theorem}

The proof of this result exploits the fact that game $\Gamma$ is a potential game to characterize the Nash equilibrium. Interestingly, we observe that non-participation by everybody, i.e., $\lambdab=(0,\ldots, 0)$, cannot be an equilibrium. Indeed, as the estimation cost diverges at $\lambdab=(0,\ldots, 0)$, every agent has a profitable deviation from this point since contributing any positive $\lambda_i$ brings the estimation cost down to a finite cost. Note, however, that this is not an artifact of the model, as it remains true if we assume that the estimation cost is bounded but large enough to exceed the privacy cost.

We observe that, as a consequence of the symmetry of the game in the homogeneous case, all the agents at equilibrium choose the same precision level, which is a function $\lambda^*=\lambda^*(n)$ of the total number of agents $n$. Then, from the discussion above, it is clear that $\lambda^*$ cannot be zero, so that all agents contribute a positive precision. %Moreover, remarkably, this equilibrium is such that all the agents choose a non-zero precision level, i.e., they decide to authorize access to their private data (potentially with added noise).

%As a consequence of the divergence of the estimation cost, which may go to infinite, there will be an incentive for at least one person to contribute something, lowering the actual estimation cost to a finite value. However, this is not an artifact of the model, as the same result remains true, if we assume the estimation cost to be bounded, but large enough to exceed the privacy cost. In particular, the fact that every agent contributes positively stems from our assumption that giving a small amount of data is cheap ($c'(0)=0$).

Due to the arbitrariness of the functions $F(\cdot)$ and $c(\cdot)$, the unique Nash equilibrium cannot be written in closed form. However, it is easily computable in practice either as the minimum of the potential function (which is convex) or as the unique solution of the following fixed point problem:
\begin{align*}
\lambda=g(n,\lambda),
\end{align*}
where function $g:\mathbb N^*\times [0,1/\sigma^2] \rightarrow [0,+\infty]$ is defined for each $\lambda\in (0,1/\sigma^2]$ and for each $n\in \mathbb N^*$ as
\begin{align*}
g(n,\lambda) = \min{\left\{\sqrt{F'\left( \frac{1}{n\lambda} \right) \frac{1}{n^2c'(\lambda)}} , 1/\sigma^2 \right\}}
\end{align*}
and is defined by continuity as $\lim_{\lambda \rightarrow 0^+}g(n,\lambda)$ for $\lambda=0$ and for each $n\in \mathbb N^*$.

%We observe that, as a consequence of the symmetry of the game in the homogeneous case, all the agents at equilibrium choose the same precision level, which is a function $\lambda^*=\lambda^*(n)$ of the total number of agents $n$. %Moreover, remarkably, this equilibrium is such that all the agents choose a non-zero precision level, i.e., they decide to authorize access to their private data (potentially with added noise).

Given the unique Nash equilibrium $\lambdab^*(n)$, the variance (in Equation \eqref{MeanVar}) of the estimate of $y_M$ obtained by the analyst at equilibrium is also a function of $n$, and given by the following expression:
\begin{align}\label{variance1stage-homo}
\sigma_M^2(\lambdab^*(n))=\frac{1}{n\lambda^*(n)}.
\end{align}
In Propositions~\ref{DecPrec} and \ref{DecEstCost} below, we derive the properties of the equilibrium precision and of the corresponding variance, when the number of agents varies.
 
\begin{proposition}\label{DecPrec}
The equilibrium precision level $\lambda^*(n)$ satisfies:
\begin{itemize}
\item[\textit{(i)}] $\lambda^*(n)$ is a non-increasing function of the number $n$ of agents, and
\item[\textit{(ii)}] $\lim_{n\to + \infty}\lambda^*(n)=0$.
\end{itemize}
\end{proposition}

Proposition~\ref{DecPrec} states that the equilibrium contribution of each agent decreases as the number of agents increases (Part~\textit{(i)}). This is a standard property in public good problems as agents choose their equilibrium contribution such that the marginal increase in the contribution cost equates the marginal decrease in the estimation cost, and the marginal effect of a single agent decreases when the number of agent increases. Proposition~\ref{DecPrec}-\textit{(ii)} shows that, in the limit when $n$ becomes very large, the contribution of each agents tends to zero (i.e., each agent adds a variance tending to infinity to her data). It is interesting to notice that, given that the equilibrium prevision level $\lambda^*(n)$ goes to zero as $n$ goes to infinity, the variance~\eqref{variance1stage-homo} cannot decrease in $1/n$ as in the standard case of the empirical mean of iid random variables of equal variance. This is because, here, the variance of each data point (or random variable) increases as the number of points increases. Yet, as the next proposition shows, the variance of the mean's estimate is still non-increasing. 

%we can notice that the variance goes to zero at a rate smaller from the standard $1/n$.

%The properties of the variance of the population estimate at equilibrium, as a function of the number of agents, are summarized in the following corollary. 
\begin{proposition}\label{DecEstCost}
The equilibrium variance of the estimate of $y_M$ satisfies:
\begin{itemize}
\item[\textit{(i)}]  $\sigma_M^2(\lambdab^*(n))$ is a non-increasing function of the number of agents $n$,  and
\item[\textit{(ii)}] $\lim_{n\to + \infty} \sigma_M^2(\lambdab^*(n))=0$.
\end{itemize}
\end{proposition}

Proposition~\ref{DecEstCost}-\textit{(i)} shows that, for the analyst, it is always better to have a larger number of agents giving data despite the fact that, when the number of agents increases, each agent gives data with smaller precision (see Proposition~\ref{DecPrec}). Proposition~\ref{DecEstCost}-\textit{(ii)} analyzes the case of a large number of agents $n$. Interestingly, when $n$ gets large, the variance goes to zero, though at a rate smaller than $1/n$ as mentioned above. (We give an expression of the rate in Section~\ref{specialcase} for special functions $F$ and $c$). 

\subsection{The Modified Estimation in the Homogeneous Case}\label{onestageetasec}

We now move to the case where the analyst can restrict the set of agents, thereby asking to access the data of only a subgroup of them, and potentially introducing a minimum precision level $\eta\in [0,1/\sigma^2]$. The final goal is to improve the estimation accuracy; formally, to estimate the mean $y_M$ with a variance strictly smaller than $\sigma_M^2(\lambdab^*(n))$. We assume that the set $S\subseteq N$ of agents who can authorize access to their data (i.e., who are solicited by the analyst) is fixed, and we analyze how the estimation varies while moving only the parameter $\eta$. This variant is modeled by the game $\Gamma(S,\eta)$ defined in Section~\ref{submodel}, where $\eta$ is now the only variable of the model. We suppose that the equilibrium precision level for the game $\Gamma(S,0)$ is s.t. $\lambda^*(s)\neq 1/\sigma^2$ since, otherwise, the estimation would already be optimal with variance $\sigma^2/s$ for $\eta = 0$.

A Nash equilibrium (in pure strategy) of the game $\Gamma(S,\eta)$ is a strategy profile $\lambdab^*\in \big[\{0\}\cup[\eta,1/\sigma^2]\big]^s$ satisfying
\begin{align}\label{NEonestageeta}
\lambda_i^* \in \argmin_{\lambda_i\in\{0\}\cup[\eta,1/\sigma^2]} J_i(\lambda_i,\lambdab_{-i}^*),\	\	\forall i\in S.
\end{align}
In the following theorem, we show that, if the analyst chooses a minimum precision level that is not ``too big'', the agents are still wishing to authorize access to their data at equilibrium. 
Recall that $S\subseteq N$ denotes the set of agents solicited by the analyst (who are the players of the game $\Gamma(S,\eta)$) and that $s=|S|$ denotes its cardinal. 

\begin{theorem}\label{theoremeta}
%Given the set of agents $S\subseteq N$, 
If $s=1$, then for any $\eta\in [0,1/\sigma^2]$, $\Gamma(S,\eta)$ has a unique Nash equilibrium $\lambda^*(s,\eta)=\max \left\{\lambda^*(1),\eta\right\}$. \\
If $s>1$, then there exists a unique parameter $\eta^*(s)\in [0,1/\sigma^2]$ s.t.:
\begin{itemize}
\item[\textit{(i)}] for any $\eta\in [0,\eta^*(s)]$, $\Gamma(S,\eta)$ has a unique Nash equilibrium $\lambdab^*(s,\eta)$, s.t., $\lambda_i^*(s,\eta)=\lambda^*(s,\eta)$ for each $i\in S$, with
\begin{align}\label{solution1stageeta2}
\lambda^*(s,\eta)=\left\{
\begin{aligned}
& \lambda^*(s) & \textrm{ if } 0\le \eta \le \lambda^*(s) \\
& \eta & \textrm{ if } \lambda^*(s) < \eta \le \eta^*(s); \\
\end{aligned}
\right.
\end{align}
\item[\textit{(ii)}] for any $\eta\in (\eta^*(s), 1/\sigma^2]$, there does not exist a Nash equilibrium $\lambdab^*(s,\eta)$ s.t. $\lambda_i^*(s,\eta)\neq 0$ for each $i\in S$.
\end{itemize}
\end{theorem}

Theorem~\ref{theoremeta} introduces the quantity $\eta^*(s)$ which, as we will see, is crucial for the analyst. 
Similarly to $\lambda^*(s)$, the value of $\eta^*(s)$ cannot be written in closed form, but it can be computed as the unique solution of the following fixed point problem:
\begin{align*}
\eta=\tilde{g}(s,\eta),
\end{align*}
where function $\tilde{g}: \mathbb N^* \times [0,1/\sigma^2]  \rightarrow [0,+\infty]$ is defined for each $\eta\in (0,1/\sigma^2]$ and for each $n\in \mathbb N^*$ as
\begin{align*}
\tilde{g}(s,\eta) = \min{\left\{\frac{F\left( \frac{1}{(s-1)\eta}\right)-F\left( \frac{1}{s\eta}\right)}{c(\eta)}\cdot \eta , 1/\sigma^2 \right\}}
\end{align*}
and is defined by continuity as $\lim_{\eta \rightarrow 0^+}\tilde{g}(s,\eta)$ in $\eta=0$ for each $n\in \mathbb N^*$.
We can also show that $\lambda^*(s) < \eta^*(s)$ for all $s$ (we obtain this result inside the proof of Theorem~\ref{opteta1stage}).

Theorem~\ref{theoremeta} characterizes the Nash equilibrium for different values of the parameter $\eta$. We observe that, as a consequence of the symmetry of the game, when $\eta \in [0,\eta^*(s)]$, the unique equilibrium of $\Gamma(S,\eta)$ is still symmetric, as it was for the unique equilibrium of the original game $\Gamma$. More specifically, if the analyst sets a minimum precision level $\eta$ smaller than the unique equilibrium precision level $\lambda^*(s)$ of game $\Gamma$, the restriction of the strategy set does not have any effect on the outcome of the game. 
On the other hand, if the analyst sets a minimum precision level $\eta$ in the interval $(\lambda^*(s), \eta^*(s)]$, all agents are still willing to participate with a precision $\eta>\lambda^*(s)$. This result matches with intuition, because even though agents' marginal costs are higher than the marginal benefits (the equilibrium choice is on the border of the strategy space $[\eta, 1/\sigma^2]$), their costs are still lower than if they choose a precision level zero. Therefore, agents do not have incentives to deviate. In the remaining range $(\eta^*(s),1/\sigma^2]$, there does not exist an equilibrium such that each agent chooses a non-zero precision level. If there exist Nash equilibria, they are such that a subset $S'\subset S$ of agents choose the non-zero precision level $\lambda^*(s',\eta)$, while the others choose zero. The possible existence of these equilibria is not relevant for our analysis. In fact, such an equilibrium would provide the same estimation that the analyst can obtain by implementing the game $\Gamma(S',\eta)$ and, as we see in the following theorem, the estimation improves by maximizing the number of agents in the game.

The previous theorem is an important stepping stone allowing us to establish the main result of this section:
%With the aim of improving the estimation, it is optimal, for the analyst, to ask the access to the private variable of all the agents whose data are contained in the data repository. Moreover, there exists an optimal minimum precision level, which allows the analyst to obtain the access to the personal data of all the agents, with an higher level of precision.
\begin{theorem}\label{opteta1stage}
The estimation variance at equilibrium is minimal for $S=N$ and $\eta = \eta^*(n)$. Moreover, we have 
\begin{equation*}
\sigma_M^2(\lambdab^*(n,\eta^*(n)))<\sigma_M^2(\lambdab^*(n)),
\end{equation*}
that is, setting a minimum precision level $\eta = \eta^*(n)$ strictly improves the estimation. 
% Asymptotically, the improvement obtained by setting the minimum precision level $\eta=\eta^*(n)$ is lower bounded by the quantity
%\begin{align}
%\lim_{x\rightarrow 0}\frac{xc'(x)}{c(x)}.
%\end{align}
\end{theorem}
Theorem~\ref{opteta1stage} shows that the analyst can indeed improve the quality of the estimation by setting a minimum precision level. It establishes that it is optimal, for the analyst, to solicit access to the private variable of all the agents whose data is contained in the data repository; and it provides the optimal minimum precision level $\eta=\eta^*(n)$ that the analyst should set to maximize the estimation precision. (Recall that $\eta^*(n)$ can be easily computed from the model's parameters by solving a fixed point problem.) Overall, Theorem~\ref{opteta1stage} provides an implementable mechanism through which the analyst can improve the quality of the data provided by each user by imposing restrictions on the variance that users can add. In the next section, we study a special case with simple functions $F(\cdot)$ and $c(\cdot)$ in order to quantify precisely the improvement achieved. 

%the optimal minimum precision level $\eta=\eta^*(n)$, which, as we have already stated, can be easily computed from the model's parameters by solving a fixed point problem. 

%With the aim of improving the estimation, it is optimal, for the analyst, to ask the access to the private variable of all the agents whose data are contained in the data repository. Moreover, there exists an optimal minimum precision level, which allows the analyst to obtain the access to the personal data of all the agents, with an higher level of precision.

\subsection{The Special Case with Monomial Privacy Costs and Linear Estimation Cost}\label{specialcase}

In this section, we illustrate the results of the previous sections on the special case where the privacy cost is monomial and the estimation cost is linear; i.e., we assume that the 
cost function in \eqref{costfunction1stage} has the form
\begin{align}\label{costfunction1stagesimple}
J_i(\lambda_i,\lambdab_{-i})= c\lambda_i^k + \sigma_M^2(\lambdab),
\end{align}
where $c\in (0, \infty)$ and $k \ge 2$ are constants. Note that, without loss of generality, in the linear estimation cost, we omit the constant factor (adding a constant to the cost does not modify the game solutions) as well as the slope factor (adding it would give an equivalent game with constant $c$ rescaled). 
For this special case, we can determine both the equilibrium precision (without a minimum precision level) and the optimal minimum precision level in closed form. We can then graphically depict how the quantities vary while moving the model parameters, and explicitly compute the estimation improvement.
%
%Note that a first preliminary analysis of the simplified model with costs as in \eqref{costfunction1stagesimple} is given in \cite{Chessa15a}. 
A preliminary analysis of the simplified model with costs as in \eqref{costfunction1stagesimple} was provided in our previous work \cite{Chessa15a}; we provide an extended analysis of this special case here thanks to the results of the previous section.
%Now, together with the results of the previous section, we are able to provide an extended analysis of this special case. 

%For illustration, we characterize explicitly the key quantities of the previous sections, under the assumption of monomial privacy costs and linear estimation cost, i.e., when the cost function in \eqref{costfunction1stage} has the form
%\begin{align}\label{costfunction1stagesimple}
%J_i(\lambda_i,\lambdab_{-i})= c\lambda_i^k + \sigma_M^2(\lambdab),
%\end{align}
%with $c\in \mathbb R$, $c>0$ and $k \ge 2$. A first preliminary analysis of such simplified model is in \cite{Chessa15a}. We are able to provide an extended analysis of it, thanks to the new results shown in the previous sections. In particular, in this special case, we can write both the equilibrium precision without minimum precision level, and the optimum minimum precision level in closed form. This makes easier to visually see how the quantities vary while moving the model parameters, and allows us to explicitly compute the estimation improvement.

In the special case of costs given by \eqref{costfunction1stagesimple}, the equilibrium precision chosen by the agents in the game $\Gamma$ simplifies to:
%When the agents play the estimation game $\Gamma$, at equilibrium they allow the access to their data, perturbed according to the following level of precision:
\begin{align}\label{lambdaspe}
\lambda^*(n)=\left\{
\begin{aligned}
& \left(\frac{1}{ckn^2}\right)^{\frac{1}{k+1}} & \textrm{ if } \left(\frac{1}{ckn^2}\right)^{\frac{1}{k+1}} \le 1/\sigma^2 \\
& 1/\sigma^2 & \textrm{ if } \left(\frac{1}{ckn^2}\right)^{\frac{1}{k+1}} > 1/\sigma^2. \\
\end{aligned}
\right.
\end{align}
As we have seen in the previous section (Theorem~\ref{opteta1stage}), it is optimal for the analyst to request access to the data of all agents in $N$. In this special case, the corresponding optimal minimum precision level becomes
\begin{align*}
\eta^*(n)=\left\{
\begin{aligned}
& \left(\frac{1}{cn(n-1)}\right)^{\frac{1}{k+1}} & \textrm{ if } \left(\frac{1}{cn(n-1)}\right)^{\frac{1}{k+1}} \le 1/\sigma^2 \\
& 1/\sigma^2 & \textrm{ if } \left(\frac{1}{cn(n-1)}\right)^{\frac{1}{k+1}} > 1/\sigma^2. \\
\end{aligned}
\right.
\end{align*}

Writing explicitly these two key quantities, we can immediately notice that, when $c$ increases, i.e., when the agents are more concerned about privacy, they choose at equilibrium a smaller precision level $\lambda^*(n)$. Further, the minimum precision level $\eta^*(n)$ proposed by the analyst becomes smaller, if the agents are more sensitive about the protection of their data. In this special case, the properties of the results for the generic case are easy to spot. For instance, we have $\lambda^*(n)<\eta^*(n) $ for each $n\in \mathbb N^*$, and both of these quantities decrease and go to zero when $n$ increases and goes to $+\infty$.

Most interestingly, the closed-form expressions that we have for this special case allow us to analyze the rate of decrease of the variance, and to quantify the improvement that can be achieved by imposing a minimum precision level. For $n$ large enough (such that both $\lambda^*(n)$ and $\eta^*(n)$ are strictly smaller than $1/\sigma^2$), the variance at equilibrium level $\lambda^*(n)$ of game $\Gamma$ is given by
$$
\sigma_M^2(\lambdab^*(n)) = \frac{1}{n\left(\frac{1}{ckn^2}\right)^{\frac{1}{k+1}}}, 
$$
while the variance at equilibrium level $\lambda^*(n,\eta^*(n))$ of game $\Gamma(N,\eta^*(n)))$ where the optimal minimum precision level is set  is given by
$$
\sigma_M^2(\lambdab^*(n,\eta^*(n))) = \frac{1}{n\left(\frac{1}{cn(n-1)}\right)^{\frac{1}{k+1}}}.
$$
Both appear to have the same rate of decrease in $n^{\frac{-k+1}{k+1}}$ which is smaller than $n^{-1}$ but becomes closer to $n^{-1}$ as $k$ tends to infinity. Intuitively, as the privacy cost becomes closer to a step function, the equilibrium precision level becomes less dependent on the number of agents so that we get closer to the case of averaging iid random variables of fixed variance. Consequently, for $n$ large enough, the improvement is given by a factor:
\begin{equation}
\label{eq.ratiovar}
\frac{\sigma_M^2(\lambdab^*(n))}{\sigma_M^2(\lambdab^*(n,\eta^*(n)))} = \left(\frac{kn}{n-1}\right)^{\frac{1}{k+1}} > 1,
\end{equation}
which asymptotically becomes constant:
\begin{equation}
\label{eq.ratiovarasympto}
\frac{\sigma_M^2(\lambdab^*(n))}{\sigma_M^2(\lambdab^*(n,\eta^*(n)))} \sim_{n\to\infty} k^{\frac{1}{k+1}}. 
\end{equation}
Interestingly, we notice that this ratio of variances (characterizing the improvement when setting the optimal minimum precision level) depends on $k$, but not on $c$. (This holds even before the asymptotic regime, as long as $n$ is large enough such that both $\lambda^*(n)$ and $\eta^*(n)$ are strictly smaller than $1/\sigma^2$.)

%What is most interesting is that the model analysis in this special case, allows us to see more in the detail how much the estimation can improve by simply adding a minimum precision level. In particular, for $n$ large enough, the variance at equilibrium level $\lambda^*(n)$ is given by
%$$
%\frac{1}{n\left(\frac{1}{ckn^2}\right)^{\frac{1}{k+1}}},
%$$
%while the variance when setting the optimal minimum precision level is given by
%$$
%\frac{1}{n\left(\frac{1}{cn(n-1)}\right)^{\frac{1}{k+1}}}.
%$$
%By comparing these two quantities as functions of $n$, we can notice that, by setting a minimum precision level, they both have the same decreasing rate, and the improvement is given by a constant factor (which, specifically, depends on $k$ and not on $c$). This can be seen more clearly by writing the ratio of the improvement
%$$
%\frac{\sigma_M^2(\lambdab^*(n))}{\sigma_M^2(\lambdab^*(n,\eta^*(n)))} = \left(\frac{kn}{n-1}\right)^{\frac{1}{k+1}} > 1, \quad (k\ge 2)
%$$
%where the parameter $c$ does not appear. Asymptotically, this quantity converges towards a constant :
%$$
%\frac{\sigma_M^2(\lambdab^*(n))}{\sigma_M^2(\lambdab^*(n,\eta^*(n)))} \sim_{n\to\infty} k^{\frac{1}{k+1}}
%$$
%which depends only on $k$.

\begin{figure}
\centering
\begin{subfigure}[t]{1.6in}
\begin{tabular}{rc}
\rotatebox{90}{\hspace{0.9cm} $k^{\frac{1}{k+1}}$} & \hspace{-0.4cm}\includegraphics[width=\textwidth]{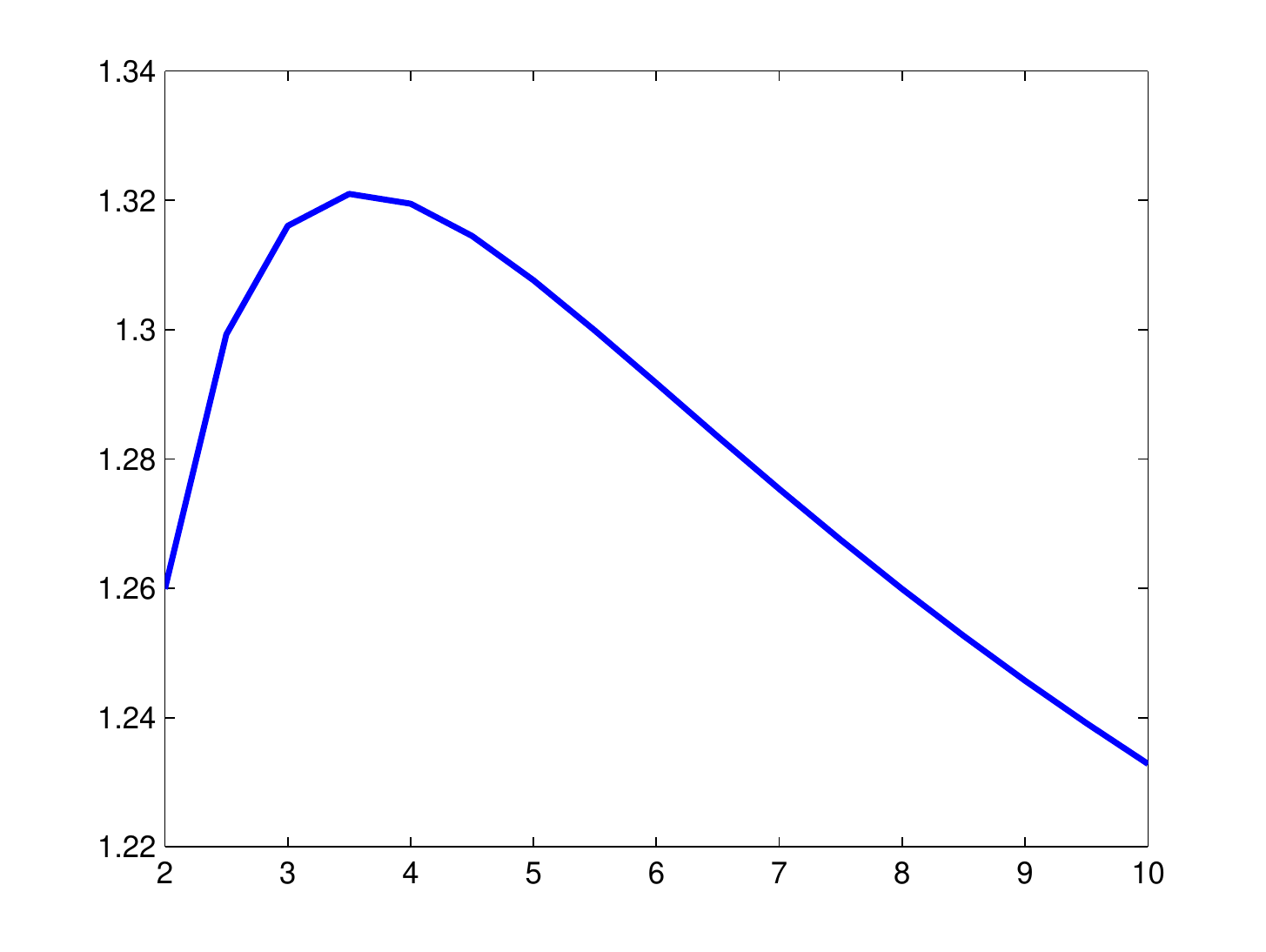} \\ 
& $k$
\end{tabular}
\end{subfigure}
\begin{subfigure}[t]{1.6in}
\begin{tabular}{c}
\hspace{0.3cm}\includegraphics[width=\textwidth]{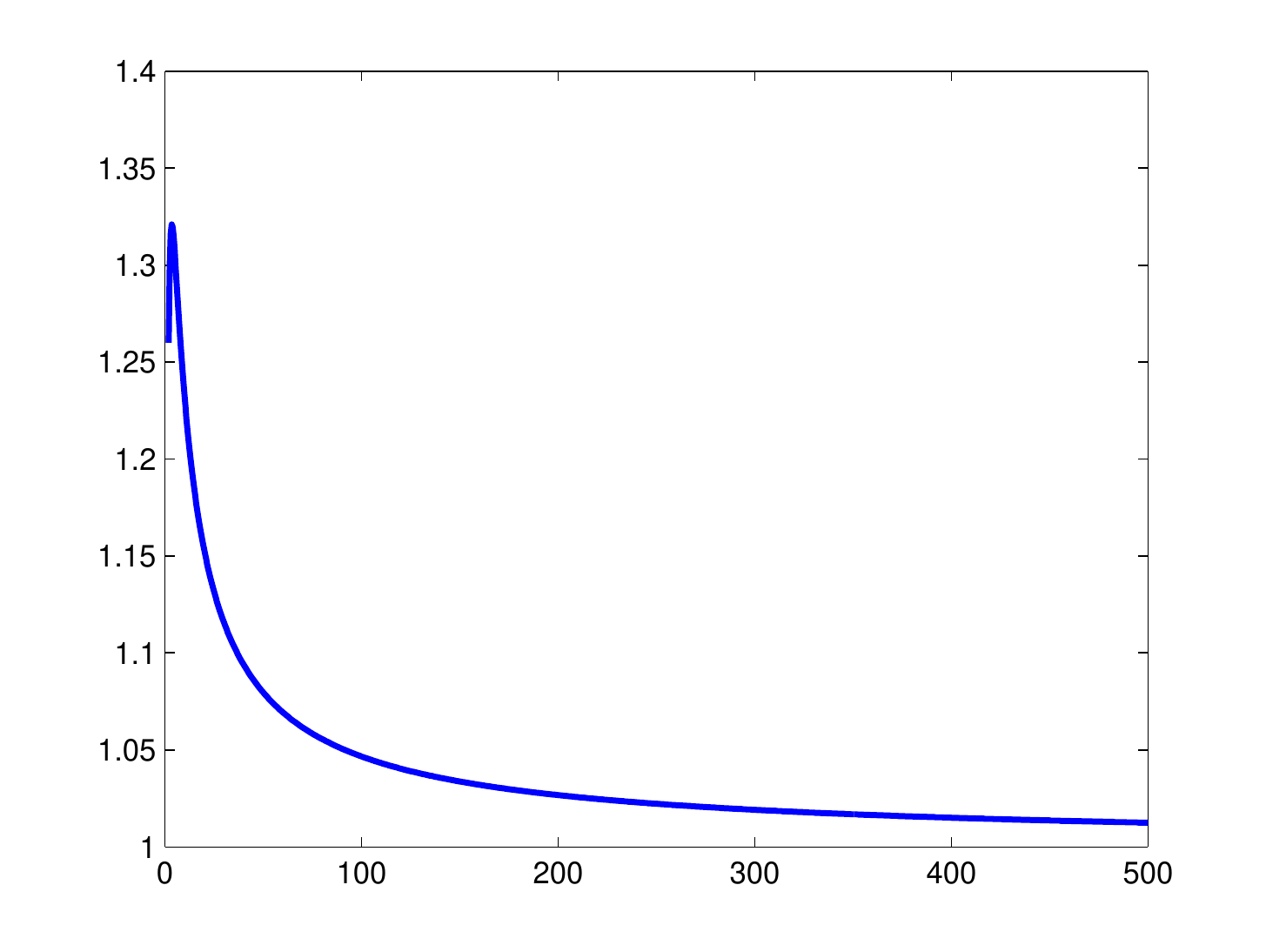} \\ 
\hspace{0.3cm}$k$
\end{tabular}
\end{subfigure}
\caption{Asymptotic improvement of the estimation choosing the optimum precision level $\eta^*$ for values of $k=2,\ldots ,10$ and for values of $k=2,\ldots ,500$.}\label{fig:figure1}
\end{figure}

Figure~\ref{fig:figure1} illustrates the asymptotic improvement ratio \eqref{eq.ratiovarasympto} for different values of $k$. We observe that it is bounded, it goes to $1$ for large $k$'s and it is in the range of $25-30\%$ improvement for values of $k$ around $2-10$. Given that the ratio \eqref{eq.ratiovar} converges towards its asymptote from above, this asymptotic improvement represents a lower bound of the improvement the analyst can achieve by implementing our mechanism with any finite number of agents $n$.

%is able to reach by implementing our model, if compared with any realistic case with a finite number of agents $n$.

%It is interesting to finally notice that, from the proof of Theorem~\ref{opteta1stage}, we also obtain that any choice of $\eta$ in $(\lambda^*(n),\bar{\eta}(n)]$ strictly improves the accuracy of the estimation of $y_M$. Note that it holds as long as $\lambda^*(n)\neq 1/\sigma^2$ as we have assumed; otherwise the estimate's accuracy is optimal even without setting a minimum precision level.

%Asymptotically, a lower bound for the improvement obtained by setting the optimal minimum precision level is given by $\lim_{x\rightarrow 0}\frac{xc'(x)}{c(x)}$.
%$$
%\frac{\sigma_M^2(\lambdab^*(n))}{\sigma_M^2(\lambdab^*(n,\bar{\eta}(n)))} = \left(\frac{kn}{n-1}\right)^{\frac{1}{k+1}} > 1, \quad (k\ge 2).
%$$
%\end{theorem}
%
%
%

%
%\subsection{The Two-stage Estimation in the Homogeneous Case with Minimum Precision Level $\eta$}\label{twostageetasec}
%
%

\section{The Heterogeneous Agent Case}\label{het}

The previous section presents an exhaustive analysis of our model in the homogeneous case, i.e., when the agents exhibit the same privacy concerns. This simplified approach enables us to derive a first set of concrete results, intuition and qualitative understanding of the model and of the minimum contribution level mechanism. The results directly apply to homogeneous populations, and can serve as a first approximation by the analyst in other cases, i.e., whenever she does not have specific information about the agents. Indeed, the results are functions only of the total number of agents, and in practice this could represent the only available detail about the agents whose data is stored in the data repository. However, not all populations are homogeneous in their privacy concerns and having more details about the different privacy concerns of the agents allows for a customized analysis. Measuring how individuals value their private information is non-trivial, but researchers have conducted direct measurement surveys \cite{Acquisti12,Spiekermann12} and various laboratory/field experiments \cite{Acquisti13b,Grossklags07} allowing for an approximate ranking of users' privacy concerns, and context-specific valuations.

With this scope, we now extend our approach to the case in which the analyst faces a heterogeneous population. In this section, we remove the restricting hypothesis of homogeneity of the agents, and we allow them to exhibit different privacy concerns. Formally, the privacy cost function of an agent $i\in N$ is equal to $c_i(\cdot)$, where all the $c_i$'s satisfy Assumption~\ref{privacyassumption}, but may be different from each other.

In order to model this situation, we follow the same approach that we used for the homogeneous case, i.e., we first analyze the situation in which the analyst implements the game $\Gamma$, without restricting the set of agents and without introducing a minimum precision level. Thereafter, we show how the analyst can improve the estimation by implementing a modified game $\Gamma(S,\eta)$. 

\subsection{The Estimation Game in the Heterogeneous Case}\label{heterogeneous.Gamma}

We start by analyzing the game $\Gamma$ where each agent's action set is $[0, 1/\sigma^2]$. 
As for the homogeneous case, also in the heterogeneous case we know that the equilibrium of the game $\Gamma$ exists and is unique because we are considering a special case of the game in \cite{ioannidis2013linear}. However, we can now characterize the equilibrium in more detail. The first result of the section, is presented in the following theorem.
\begin{theorem}\label{nonhomtheorem}
%Suppose that the agents are ordered s.t. $c_1'(\lambda) \le ... \le c_n'(\lambda)$, for each $\lambda\in [0,1/\sigma^2]$. 
Assume that the privacy costs satisfy $c_1'(\lambda) \le \cdots \le c_n'(\lambda)$, for all $\lambda\in [0,1/\sigma^2]$. 
Then, game $\Gamma$ has a unique Nash equilibrium $\lambdab^*$ s.t., $0 < \lambda^*_n \le \cdots \le \lambda^*_1$.
\end{theorem}

Theorem~\ref{nonhomtheorem} assumes that the agents can be ordered in such a way that, for any precision level $\lambda \in [0,1/\sigma^2]$, an agent choosing precision $\lambda$ has higher marginal privacy cost (and hence higher privacy cost since $c_i(0)=0$ for all agents) than the previous agents if they choose the same precision level. This may require some re-ordering from the initial ordering, which comes without loss of generality. We believe that this assumption will often be reasonable in practice since agents who are more reluctant to increase the precision of their revealed data from a small precision (i.e., have higher marginal privacy cost for a small $\lambda$) will likely be more reluctant to increase the precision of their revealed data from a large precision (i.e., have higher marginal cost for a large $\lambda$ too).

The proof of Theorem~\ref{nonhomtheorem} exploits the potential nature of the game to characterize the Nash equilibrium. The unique Nash equilibrium, which cannot be written in closed form, can be easily computed as the minimum of the (convex) potential function of the game $\Gamma$, which is the function $\Phi:[0,1/\sigma^2]^n \rightarrow \bar{\mathbb R}_+$, s.t., for each $\lambdab\in [0,1/\sigma^2]^n$,
\begin{align}\label{potentialfunction1stagemaintext}
\Phi(\lambdab) = \sum_{j\in N}c_j(\lambda_j) + f(\lambdab).
\end{align}

We observe that, in the heterogeneous case, due to the asymmetry of the model, we no longer have a symmetric equilibrium. Moreover, the equilibrium strategy cannot be written as a function of the total number of agents $n$, as it depends on their privacy cost functions. We will use the notation $\lambdab^*=\lambdab^*(N)$ to denote that the equilibrium depends on the specific identity of the agents in the set of agents $N$. As expected, at equilibrium, agents with higher privacy concerns select lower precisions and, as for the homogeneous case, no agent decides to deny the access to her data. The fact that every agent contributes positively at Nash equilibrium stems from our assumption that giving a small amount of data implies very little cost since the marginal cost at zero is zero ($c'(0)=0$). (Note, though, that some agents may contribute arbitrarily close to zero.) This assumption, although realistic, is not strictly necessary; but it greatly simplifies the presentation of our model and results.

%Note though, that in the heterogeneous case, some agents may contribute arbitrarily close to zero. This assumption is not strictly necessary but it greatly increases the elegance of our model and results. Moreover, we believe that this is a realistic assumption. In the revision, we will discuss the cost model in more detail, and further justify the robustness of our results to our assumptions.

%As a consequence of the divergence of the estimation cost, which may go to infinite, there will be an incentive for at least one person to contribute something, lowering the actual estimation cost to a finite value. However, this is not an artifact of the model, as the same result remains true, if we assume the estimation cost to be bounded, but large enough to exceed the privacy cost. In particular, the fact that every agent contributes positively stems from our assumption that giving a small amount of data is cheap ($c'(0)=0$).

Given the unique Nash equilibrium $\lambdab^*(N)$, the variance \eqref{MeanVar} of the estimate of $y_M$ obtained by the analyst at equilibrium is given by the following expression:
\begin{align}\label{variance1stage}
\sigma_M^2(\lambdab^*(N))=\frac{1}{\sum_{j\in N}\lambda_j^*(N)}.
\end{align}
Even if the equilibrium precisions chosen by the agents (and the corresponding variances) are not functions of only $n$, we can still generalize Propositions \ref{DecPrec} and \ref{DecEstCost} to the heterogeneous case. In Propositions \ref{DecPrecHet} and \ref{DecEstCostHet}, we analyze how the equilibrium precision and the variance of the estimate at equilibrium vary when a new additional agent enters the game. Note that the following two propositions do not use the ordering assumption of Theorem~\ref{nonhomtheorem}. 

\begin{proposition}\label{DecPrecHet}
Given the game $\Gamma$, suppose that an additional $(n+1)$-th agent enters the game, and denote by $\lambdab^*(N\cup\{n+1\})$ the new equilibrium precision level. Then, for each $i\in N$, $\lambda_i^*(N\cup\{n+1\}) \le \lambda_i^*(N)$.
\end{proposition}

Proposition~\ref{DecPrecHet} states that the equilibrium contribution of each agent decreases, as soon as a new agent enters the game. 

\begin{proposition}\label{DecEstCostHet}
Given the game $\Gamma$, suppose that an additional $(n+1)$-th agent enters the game. Then, $\sigma_M^2(\lambdab^*(N\cup\{n+1\})) \le \sigma_M^2(\lambdab^*(N))$.
\end{proposition}

Proposition~\ref{DecEstCostHet} shows that, for the analyst, it is always better to let new agents enter the game despite the fact that, doing so, each other agent is giving data with a lower precision. Surprisingly, this is true even if the agent who enters has higher privacy concerns than any other agent in the game, and then would accordingly contribute the lowest quality data. %Adding bad quality data is not disadvantageous for the analyst.

\subsection{The Modified Estimation in the Heterogeneous Case}\label{onestageetasecHet}

We now move to the case where the analyst can restrict the set of agents by introducing a minimum precision level $\eta\in[0,1/\sigma^2]$. Again, her final goal is to improve the estimation accuracy. We consider at first the set of agents $S\subseteq N$ to be fixed, and we analyze how the estimation varies while moving only the parameter $\eta$. This variant is modeled by the game $\Gamma(S,\eta)$ defined in Section~\ref{submodel}, where $\eta$ is now the only variable of the model. We denote by $\lambdab^*(S)$ the equilibrium precision level for the game $\Gamma(S,0)$, and we suppose that it is such that there exists at least one agent $i\in S$ s.t. $\lambda_i^*(S)\neq 1/\sigma^2$; otherwise the estimation is already optimal with variance $\sigma^2/s$ for $\eta = 0$.

The next result extends Theorem~\ref{theoremeta} to the heterogeneous case. We show that, if the analyst selects a minimum precision level which is not ``too high'', at equilibrium, all the agents (even the most concerned about privacy) are still willing to authorize access to their data (with perturbation).

\begin{theorem}\label{theoremetaHet}
As in Theorem~\ref{nonhomtheorem}, assume that the privacy costs satisfy $c_1'(\lambda) \le \cdots \le c_n'(\lambda)$, for each $\lambda\in [0,1/\sigma^2]$. Given the set of agents $S\subseteq N$, with cardinality $s\ge1$:
\begin{itemize}
\item[\textit{(i)}] if $s=1$, then for any $\eta\in [0,1/\sigma^2]$, $\Gamma(S,\eta)$ has a unique Nash equilibrium $\lambda^*_1(S,\eta)=\max \left\{\lambda_1^*(S),\eta\right\}$;
\item[\textit{(ii)}] if $s>1$, then there exists a parameter $\eta^*(S)\in (\lambda^*(S),1/\sigma^2]$ such that, for any $\eta\in [0,\eta^*(S)]$, $\Gamma(S,\eta)$ has a unique Nash equilibrium $\lambdab^*(S,\eta)$ with $\lambda^*_i(S,\eta)>0$ for all $i\in S$.
\end{itemize}
\end{theorem}

Theorem~\ref{theoremetaHet} introduces a parameter $\eta^*(S)$ such that if the analyst sets a minimum precision level in $[0, \eta^*(S)]$, even the most privacy-concerned of the agents in $S$ does not have an incentive to deviate to a zero precision level. As the theorem is stated, $\eta^*(S)$ is not unique (any value smaller than a valid $\eta^*(S)$ but still larger than $\lambda^*(S)$ will be suitable). However, let $\eta^*(S)$ be s.t.
\begin{align}\label{eq9maintext}
& c_n(\lambda_n^*(S,\eta^*(S)))=\\
\nonumber & F\left( \frac{1}{\sum_{j\in N,j\neq n}\lambda_j^*(S,\eta^*(S))}\right) - F\left( \frac{1}{\sum_{j\in N}\lambda_j^*(S,\eta^*(S))}\right),
\end{align}
where $\lambdab^*(S,\eta^*(S))$ is the local minimum of the potential function $\Phi$ defined as in \eqref{potentialfunction1stagemaintext}, but on the domain $[\eta^*(S), 1/\sigma^2]^s$. We can prove that this $\eta^*(S)$ is unique, that it satisfies Theorem~\ref{theoremetaHet}-\textit{(ii)} and we conjecture that this definition gives the largest possible parameter satisfying Theorem~\ref{theoremetaHet}-\textit{(ii)}. 

%As we can see in detail in the proof of Theorem~\ref{theoremetaHet}, in Appendix~\ref{theoremetaHetproof}, the parameter $\eta^*(S)$ is s.t. even the most privacy concerned between the agents in $S$, does not have incentives to deviate to a zero precision level. This result allows us to establish the main results of this section. It is possible, for the analyst, to strictly improve the estimation of the mean $y_M$, simply setting a minimum precision level and asking the authorization to access the data of all the agents in $N$.

The result of Theorem~\ref{theoremetaHet} allows us to establish the main result of this section: 
\begin{theorem}\label{opteta1stageHet}
As in Theorem~\ref{nonhomtheorem}, assume that the privacy costs satisfy $c_1'(\lambda) \le \cdots \le c_n'(\lambda)$, for each $\lambda\in [0,1/\sigma^2]$. Let $\eta^*(N)$ be as in Theorem~\ref{theoremetaHet}-\textit{(ii)} for $S=N$. The analyst can improve the estimation by implementing the game $\Gamma(N,\eta^*(N))$ with minimum precision level $\eta^*(N)$, i.e., 
$$
\sigma_M^2(\lambdab^*(N,\eta^*(N))) < \sigma_M^2(\lambdab^*(N)).
$$
\end{theorem}

Theorem~\ref{opteta1stageHet} shows that the analyst can improve the precision of the estimation of the mean $y_M$ simply by setting a minimum precision level and soliciting access to the data from all the agents in $N$. This is true for any minimum precision level $\eta^*(N)$ such that Theorem~\ref{theoremetaHet}-\textit{(ii)} is satisfied and shows that, even in the heterogeneous case, it is possible to strictly improve the estimation by applying the minimum precision level mechanism. Here too, however, we conjecture that the parameter $\eta^*(N)$ solving \eqref{eq9maintext} yields the highest possible improvement. 

%It is possible, for the analyst, to strictly improve the estimation of the mean $y_M$, simply setting a minimum precision level and asking the authorization to access the data of all the agents in $N$.

\subsection{The Special Heterogeneous Case with Monomial Privacy Costs and Linear Estimation Cost}\label{specialcaseHet}

As for the homogeneous case, we now illustrate the results of the previous sections on the heterogeneous model in the special case of monomial privacy cost and linear estimation cost. 
%Also for the heterogeneous case, for illustration, we characterize explicitly the results under the assumption of monomial privacy costs and linear estimation cost. 
In this simplified model, the cost function in \eqref{costfunction1stage} has the form
\begin{align}\label{costfunction1stagesimpleHet}
J_i(\lambda_i,\lambdab_{-i})= c_i\lambda_i^k + \sigma_M^2(\lambdab),
\end{align}
with $c_i\in (0, \infty)$ for each $i\in N$ and $k \ge 2$. The assumption of Theorem~\ref{nonhomtheorem} that agents can be ordered s.t. $c_1'(\lambda) \le ... \le c_n'(\lambda)$ for each $\lambda\in [0,1/\sigma^2]$, translates now to requiring that $0 <  c_1 \le ... \le c_n$ (which, in the case of monomial costs, is completely without loss of generality).

Even with such a simplified model, having heterogeneous agents does not allow us to write the key quantity in closed form as we did in the simplified homogeneous model in Section~\ref{specialcase}. However, it is still possible to provide clearer expressions and to quantify the variance improvement by setting a minimum precision level.

When the agents play the estimation game $\Gamma$, at equilibrium they choose a precision level that, if interior, can be written as
$$
\lambda_i^*(N)=\left( \frac{1}{c_i k \left( \sum_{j\in N}\lambda_j^*(N)\right)^2}\right)^{\frac{1}{k-1}}.
$$
The analyst can improve the estimation by setting a minimum precision level $\eta^*(N)$. In this simplified case, it takes the form
\begin{align*}
& \eta^*(N) = \\
& \!\!\!\! \left(\frac{1}{\!c_n\! \left(\sum_{j\in N}\lambda_j^*(\eta^*(N))\right)\! \left(\sum_{j\in N\setminus \{n\}}\lambda_j^*(\eta^*(N))\right)}\right)^{\frac{1}{k-1}}.
\end{align*}

Note that the two expressions above are in the form of fixed-point equations. It is interesting to note that when $k>c_n/c_1$ though, i.e., when the privacy cost of the agents are not too dispersed, this minimum precision level can be written in closed form as
\begin{align}
\eta^*(N)=\left(\frac{1}{c_n n (n-1)}\right)^{\frac{1}{k+1}}.
\end{align}
It is then equal to the optimal precision level, when all the agents have the same privacy cost as the most privacy-concerned individual.

\begin{figure}[h!]
\centering
\begin{tabular}{rc}
\rotatebox{90}{\hspace{0.7cm} Ratio of variances} & \hspace{-0.0cm}\includegraphics[width=2.2in]{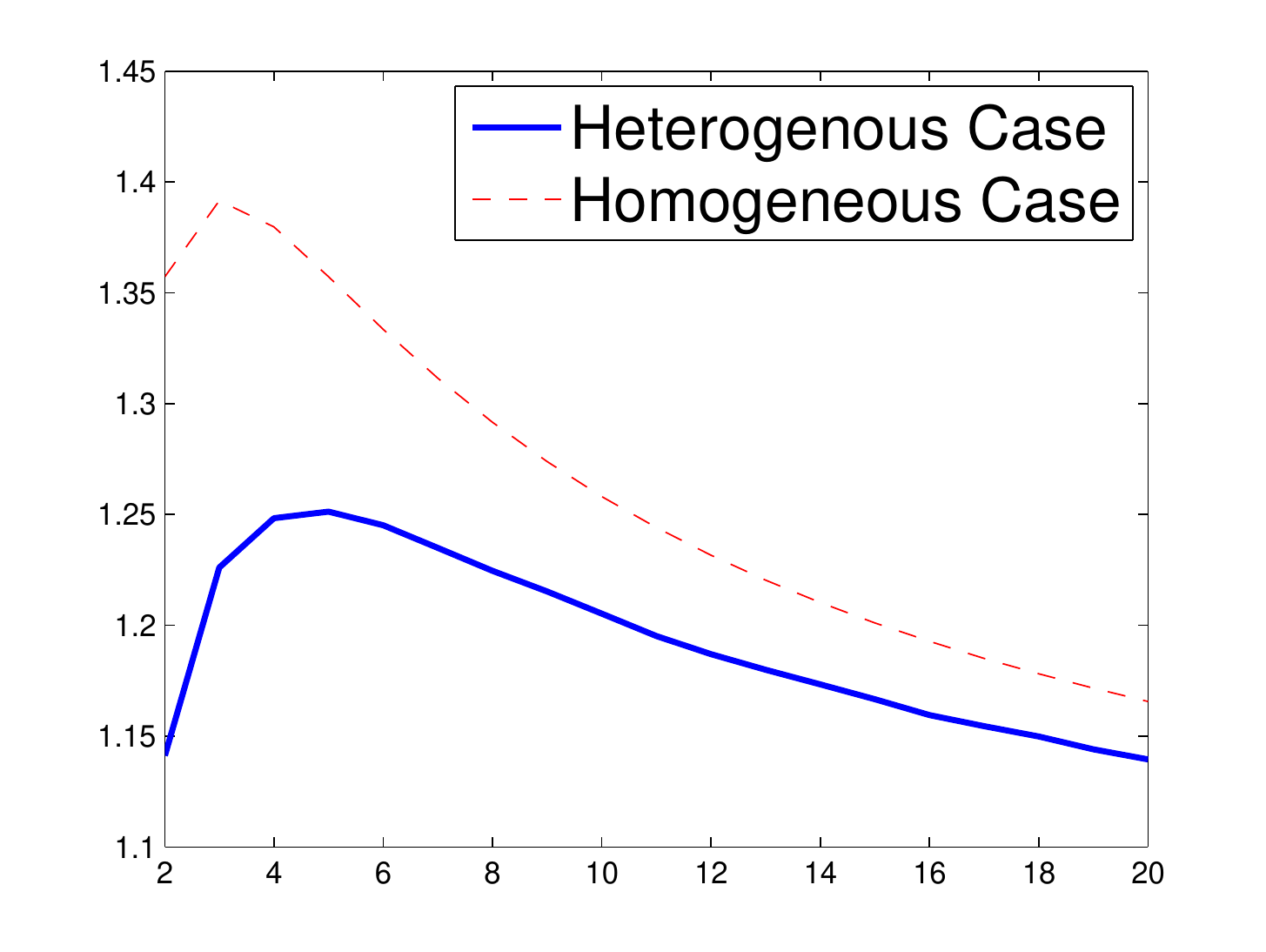} \\ 
& $k$
\end{tabular}
\caption{Improvement of the estimation in $\Gamma(\eta)$ in the heterogeneous case choosing the optimum precision level $\eta^*(N)$, compared to the homogeneous case choosing the optimum precision level $\eta^*(n)$; for values of $k=2,\ldots ,20$. In this example, $\textbf{c}= (1,1.5,2,2.5,3)$, $1/\sigma^2=2$.}\label{fig:graphfinal}
\end{figure}

Figure~\ref{fig:graphfinal} illustrates on an example the estimation improvement in the heterogeneous case when choosing $\eta^*(N)$ as above (which we conjectured is the optimal choice). We compare it with the improvement in the analogous homogeneous case when choosing the optimal $\eta^*(n)$ (see Theorem \ref{opteta1stage} which does not depend on $c$).

\section{Extensions of the Model}\label{extensions}
In this section, we extend our model in two directions. In Section~\ref{twostage}, we propose an alternative modified estimation game, and we compare it with the one proposed in Section~\ref{submodel}. The main difference with the previous one is that it is a two-stage game. 
In Section~\ref{costs}, we add an  important variable to our model by introducing a per-agent cost of collecting data. 
%We suppose that the analyst collecting the data has to face some per agent costs. 
%As a consequence, we will observe that it is not convenient anymore for the analyst to have an arbitrarily large set of agents. 
%In Section~\ref{stability}, we propose a stability analysis of the modified game $\Gamma(N,\eta^*(n))$, in case the analyst does not know exactly the number of agents whose data is in the repository.
Both proposed extensions are included to derive qualitative insights about the practical applicability of the model, however, we defer an in-depth analysis to future work.

\subsection{The Modified Two-Stage Game}\label{twostage}

In $\Gamma(N,\eta)$, both the decision to authorize the access (or to deny it) and the selection of a precision level (in case of authorization) are simultaneous. This variant captures cases in a realistic fashion where the analyst requests access to data already present in a repository. In different applications, however, the analyst may first recruit participants that commit to provide private data with a minimum precision; and only in a second stage (for example, as soon as the data becomes available), these agents would be asked to disclose their information. This scenario applies, for example, to medical research studies or consumer decisions, and it motivates the study of a model where agents first decide to participate or not, and only then decide on the precision of the data released. Another motivation to study such a model is that it could lead to a higher estimation accuracy, in which case the analyst would want to implement it even if it does not naturally arise from the application at stake.
 
In this section, we investigate this extension of our original model in the simplified case with homogeneous monomial privacy costs and linear estimation cost as it is sufficient to understand and illustrate the qualitative differences between the two models. We leave the development of the more general model to future work. We also point out the possibility, for future work, of a similar extension, in which the agents asynchronously make decisions on whether or not to share their data (i.e., they make their sharing decisions based on actions taken by agents who were contacted earlier by the analyst). However, in absence of observability of the contribution decisions (as it is often the case in the medical domain due to confidentiality restrictions) even asynchronous decision-making can be approximated well with a simultaneous move model.

%To address these different problems, and to investigate which one of the two scenarios allows for a more efficient collection of data, we compare the game $\Gamma(N,\eta)$ with another variant, that we define next. 
%We conduct our analysis for the special case of monomial privacy costs and linear estimation cost of Section~\ref{specialcase}.

To investigate our variant of the model and to compare its outcome with the one of the game $\Gamma(N,\eta)$, we define a two-stage variant of the game. 
We assume that the agents are initially informed of the minimum precision level $\eta$. In a first stage, they have to decide if they want to deny access to their data, and exit the game, or if they wish to accept to authorize access. The set of agents who accepted to participate is revealed to all agents. In a second stage, the agents who decided to participate choose their precision in the imposed range $[\eta, 1/\sigma^2]$. Formally, this situation is modeled through the following two-stage game $\Gamma^2(\eta)$:
\begin{itemize}
\item[\textit{(i)}] In the {\em first stage}, the agents make a binary choice $p_i\in \{0,1\}$
\begin{align*}
\forall i\in N\	\	p_i=\left\{
\begin{aligned}
& 0 \textrm{ if }i\textrm{ denies the access} \\
& 1 \textrm{ if }i\textrm{ accepts to authorize}.
\end{aligned}
\right.
\end{align*}
We denote by $\pb\in \{0,1\}^n$ a strategy profile, $P=\{i\in N: p_i=1\}$ the set of agents who accept, and $p=|P|$ its cardinality.
\item[\textit{(ii)}] In the {\em second stage}, given $\pb\in \{0,1\}^n$, the agents play a game $\Gamma^P(\eta)=\left\langle N,[\eta,1/\sigma^2]^p\times\{0\}^{n-p}, (J_i)_{i\in N} \right\rangle$, where each agent $i\in P$ has strategy space $[\eta,1/\sigma^2]$ whereas each agent $i\in N \setminus P$ has strategy space $\{0\}$ (i.e., the agents in $N\setminus P$ can only choose $\lambda_i=0$, which, we reiterate, is equivalent to no participation). 
\end{itemize}

We have already seen that the analyst can improve the estimation by modifying the original game, and that the optimal choice, in that previous setting (in the homogeneous case), is to implement the game $\Gamma(N,\eta^*(n))$. We now investigate whether the analyst can improve the estimation even more, while implementing the game $\Gamma^2(\eta)$ for an optimal choice of the minimum precision level $\eta$.

The games $\Gamma(N,\eta)$ and $\Gamma^2(\eta)$ differ in the information available to agents when choosing their precision (observe for instance that $\Gamma(N,0)=\Gamma$, while $\Gamma^2(0)\neq \Gamma$). In $\Gamma(N,\eta)$, both the decision to authorize the access or to deny it and the decision of the precision (in case of authorization) are simultaneous. In contrast, in $\Gamma^2(\eta)$, the set of agents who will authorize with precision of at least $\eta$ is known at the time of choosing the exact precision. 

As we did for the previous games, we study $\Gamma^2(\eta)$ as a \textit{complete information game} between the agents, i.e., we assume that the set of agents, the action sets (in particular, when present, the value of the parameter $\eta$) and the costs are known by all the agents. 

We study the pure strategy Nash equilibria of the game using backward induction. Given $\pb\in \{0,1\}^n$ the outcome of the first stage, a Nash equilibrium for the second stage is a strategy profile $\lambdab^*\in [\eta,1/\sigma^2]^p\times\{0\}^{n-p}$ s.t., for each $i\in N\setminus P$, $\lambda_i^*=0$, while for each $i\in P$, $\lambda^*_i$ is s.t.
\begin{align}\label{NE2ndstageeta}
\lambda_i^* \in \argmin_{\lambda_i\in [\eta,1/\sigma^2]} J_i(\lambda_i,\lambdab_{-i}^*).
\end{align}

If for each $\pb\in \{0,1\}^n$ the second stage game has a unique solution $\lambdab^*(\pb,\eta)$ (as we will see, it is the case in our model), the choice that the agents make in the first stage determines univocally the outcome of the two-stage game. Then, $\Gamma^2(\eta)$ reduces to a one-stage game $\left\langle N,\{0,1\}^n, (J_i^1)_{i\in N} \right\rangle$, where the cost function $J_i^1:\{0,1\}^n\to \bar{\mathbb R}$, for each $\pb\in \{0,1\}$, is given for all $i\in N$ by 
\begin{align}\label{costfunction1ststage}
J_i^1(\pb) = J_i(\lambdab^*(\pb,\eta)) = c\lambda_i^*(\pb,\eta)^k + \sigma^2_M(\lambdab^*(\pb,\eta)).
\end{align}

Then, an equilibrium of the game $\Gamma^2(\eta)$ is a strategy profile $\pb^*\in \{0,1\}$ s.t.
\begin{align}\label{NE2ndstageetap}
p_i^* \in \argmin_{p_i\in \{0,1\}} J_i^1(p_i,\pb_{-i}^*),\	\	\forall i\in N.
\end{align}

We apply backward induction, by starting to analyze the second stage game. In the following lemma, we show the existence and the uniqueness of a Nash equilibrium for the game $\Gamma^P(\eta)$ when $\pb\neq (0,\ldots ,0)$.

\begin{lemma}\label{theorem2stage2nd}
For each $\pb\in \{0,1\}^n\setminus \{(0,\ldots ,0)\}$, the game $\Gamma^P(\eta)$ has a unique Nash equilibrium $\lambdab^*(\pb,\eta)$, s.t., $\lambda_i^*(\pb,\eta)=\lambda^*(p,\eta)$ for each $i\in P$, with
\begin{align}\label{solution2ndstage}
\lambda^*(p,\eta)=\left\{
\begin{array}{ll}
\lambda^*(p) & \textrm{if }  0\le \eta \le \lambda^*(p) \\
\eta & \textrm{if } \lambda^*(p)< \eta \le 1/\sigma^2,
\end{array}
\right.
\end{align}
(where $\lambda^*(p)$ is defined as in \eqref{lambdaspe}) and $\lambda^*_i(\pb,\eta)=0$, for each $i\in N\setminus P$.
\end{lemma}

We observe again that the equilibrium of the game restricted to agents in $P$ is symmetric (i.e., each participating agent chooses the same precision level at equilibrium). We call $\lambda^*(p,\eta)$ the equilibrium precision for agents in $P$ to emphasize the dependence on the cardinality $p$ of the set $P$ and on the parameter $\eta$. In fact, due to the symmetry, the optimal choice for an agent who decided to participate depends only on the number of agents who made the same choice as her in the first stage and not on the identity of these agents. Further, given $P$ and $\eta$, the equilibrium of $\Gamma^P(\eta)$ is the same as the equilibrium of $\Gamma(N,\eta)$ given in Theorem~\ref{theoremeta}, when replacing $n$ by $p$. The only difference is that, even for large $\eta$ the agents in $P$ will choose precision level $\eta$ in $\Gamma^P(\eta)$ since they are committed to participate with precision of at least $\eta$. Consequently, the equilibrium of $\Gamma^P(\eta)$ always exists and it is s.t. each agent choosing a non-zero precision level.

As the second stage game has always a unique solution, we can apply backward induction, and the two-stage game $\Gamma^2(\eta)$ reduces to a one-stage game. The following lemma establishes the existence and uniqueness of its equilibrium for a minimum precision level. 
 
\begin{lemma}\label{theorem2stage1st}
For any $\eta\in [\lambda^*(n-1),\eta^*(n)]$, the two-stage game $\Gamma^2(\eta)$ has a unique equilibrium given by $\pb^*=(1,\ldots ,1)$.
\end{lemma}

Lemma~\ref{theorem2stage1st} states that, if the analyst chooses a minimum precision level in the range $[\lambda^*(n-1),\eta^*(n)]$ and implements the two-stage game $\Gamma^2(\eta)$, then each agent will participate at equilibrium. The equilibrium contributions, given by Lemma~\ref{theorem2stage2nd}, equal $\eta$ for each agent since $\eta\ge \lambda^*(n-1)\ge\lambda^*(n)$. For $\eta$ in the range $[\lambda^*(n-1),\eta^*(n)]$, the outcome of the two-stage game $\Gamma^2(\eta)$ is therefore the same as for the one-stage game $\Gamma(N,\eta)$. This is not the case, however, for other ranges of parameters. In particular, for $\eta<\lambda^*(n-1)$, all agents participate in $\Gamma(N,\eta)$ whereas they may not participate in $\Gamma^2(\eta)$. This is because, in $\Gamma^2(\eta)$, agents react in the second stage to the participation decisions of the first stage (typically if an agent chooses not to participate, the others increase their precisions in the second stage). As a consequence, agents can strategically choose their participation in the first stage to influence the precisions chosen in the second stage. Analysis of the existence and uniqueness of the Nash equilibrium in $\Gamma^2(\eta)$ in ranges of $\eta$ outside $[\lambda^*(n-1),\eta^*(n)]$ is therefore more intricate. Nevertheless, we can establish our main result, namely that choosing $\eta=\eta^*(n)$ yields an optimal estimation variance for the analyst:
\begin{theorem}\label{opteta2stage}
For the game $\Gamma^2(\eta)$, with $\eta\in [0,1/\sigma^2]$, the estimate's variance at equilibrium $\sigma_M^2(\lambdab^*(\pb^*,\eta))$ is minimized for $\eta = \eta^*(n)$. The improvement obtained by setting the minimum precision level $\eta = \eta^*(n)$ is characterized, for $n$ large enough, by the ratio
\begin{align*}
\frac{\sigma_M^2(\lambdab^*(n))}{\sigma_M^2(\lambdab^*(\pb^*,\eta^*(n)))} = \left(\frac{kn}{n-1}\right)^{\frac{1}{k+1}} > 1, \quad (k\ge 2).
\end{align*}
\end{theorem}

Theorem \ref{opteta2stage} shows that the optimal $\eta$ is the same for the one-stage game $\Gamma(N,\eta)$ and the two-stage game $\Gamma^2(\eta)$, and both yield the same improvement for the analyst. As such, the discussion given in Section~\ref{onestageetasec} about the asymptotic behavior of this gain still holds.
However, as mentioned, the two games $\Gamma(\eta)$ and $\Gamma^2(\eta)$ are not equivalent for each choice of the parameter $\eta$. In particular, we can infer from the proof of Theorem \ref{opteta2stage} that there is still a range of minimum precision levels for which the estimation is strictly improved, but this range is smaller than it was for $\Gamma(N,\eta)$.

\subsection{The Estimation Game in the Presence of Per-Agent Costs}\label{costs}

%We propose here some preliminary results about a possible extension of this work. 
In this section, we propose an extension of our model to include the cost of collecting data. 
Indeed, in Section~\ref{mod} and throughout this paper, we assumed that data is collected at no cost, and that the analyst aims at minimizing the variance of the mean estimation. The absence of per-agent cost (to solicit contributions) is a standard assumption in most of the public good literature. However, it could limit the appeal of our model in some applications.
Here, we present preliminary results with arbitrary per-agent cost, restricted to the homogeneous case. We then introduce a simplified case with linear per-agent cost, to illustrate the qualitative difference to the zero per-agent cost case, in particular, the existence of an optimal number of agents $n$. The derivation of the optimal $n$ would be slightly different when assuming, for example, a concave cost function. This is left as a possible future work suggestion (see Section~\ref{conclusions}).

When facing a per-agent cost, we can no longer rely on the fact that the analyst will always prefer to have the largest possible set of agents. Rather, she has to select the optimal subset of agents to include in the game. In the homogeneous case, selecting the optimal subset of agents reduces to selecting the optimal number of agents $n^*\in N$. To address this problem, we assume that, instead of aiming at minimizing the variance, the analyst aims at minimizing a cost function $J_A: \mathbb N^* \rightarrow \mathbb R$ defined as
\begin{align}
J_A(n) = f(\etab^*(n)) + C n,
\end{align}
where $f$ is the estimation cost defined in Section~\ref{mod}, while $C$ represents the \textit{per-agent cost} of collecting personal data. We assume that the estimation cost is evaluated at equilibrium, when the analyst chooses the optimal minimum precision level. In fact, for a fixed $n$, $\etab^*(n)$ provides the minimum variance and, consequently, the minimum estimation cost. The problem of the analyst now reduces to setting an optimal number of agents $n^*$.

\begin{theorem}\label{costtheorem}
The function $J_A(n)$ has a minimum in $\mathbb N^*$. The optimal $n^*$ is given by $n^* = \max \{m\in N^*|c(\eta^*(m))\ge C\}$, if this set is non-empty, and by $1$ otherwise.
\end{theorem}
%This result provides the analyst with practical insights on how to optimize the number of agents which she would solicit data from in the presence of per-agent cost. 

Theorem~\ref{costtheorem} shows how the analyst can optimize the balance between the minimization of the estimation cost and the per-agent recruitment cost. In this situation, it is typically not optimal anymore to contact as many agents as possible. Of course, if the theoretically determined optimal number of agents equals or exceeds the size of the potential participant pool ($n^* \geq n$), then the analyst will contact all available agents. As $c(\eta^*(m))$ is non-increasing in $m$, $n^*$ can be easily computed by the analyst, for example by implementing a bisection method on $[1,n]$, where $n$ is the total number of agents whose data is contained in the repository.

\section{Concluding remarks}\label{conclusions}

In this paper, we investigate the problem of estimating population averages from data provided by privacy-sensitive users. We assume that users can perturb their data before revealing it (e.g., by adding zero-mean noise) in order to protect their privacy. Users, however, benefit from a more accurate population estimate. Therefore, each user strategically selects the precision of her revealed data to balance her privacy cost and the cost incurred by a lower precision of the population estimate. We find that the resulting game has a unique Nash equilibrium and carefully study its properties.

We further prove that the analyst can increase the population estimate's accuracy simply by imposing a minimum precision level for the data which users can reveal (e.g., by restricting the variance of the noise users can add). %We find that for a fixed population of data-providing users, increasing the precision of each data point improves the population estimate's precision. 
The surprising and important aspect of this result is that the scheme remains incentive-compatible, i.e., users are willing to provide data with a higher precision rather than dropping out. 
We also show how to tune the minimum precision level the analyst should set in order to optimize the population estimate's accuracy. In our numerical simulations, the maximum improvement of the population estimate's accuracy is in the order of $20-40\%$.

Our model treats the population estimate's accuracy as a public good (e.g., if one agent increases the precision of the data she gives, it benefits all users). Then, our results offer a novel method to increase the provision of a public good above voluntary contributions, simply by restricting the agents' strategy spaces. This method is attractive through its simplicity compared for instance to other schemes that involve monetary transfers, and could find application in other public good problem domains. 

The results are derived for arbitrary functions for the privacy cost experienced by each user and for the estimation cost (satisfying relatively mild assumptions). This increases the robustness of our main results and allows for application to various situations. Further, we study the cases of homogeneous and heterogeneous agents. Indeed, for practical utilization of our work it is important to be able to accommodate different types of privacy preferences as evidenced by the literature on the value of privacy (which includes direct measurement surveys \cite{Acquisti12,Spiekermann12} and laboratory/field experiments \cite{Acquisti13b,Grossklags07}).

We also consider extensions of our basic model such as variations in the structure of decision-making. Introducing a two-stage structure impacts the available information to individuals, i.e., whether or not the set of contributing agents is determined before agents choose their precision levels. Surprisingly, we find that providing this information to users can never improve the estimation's accuracy. %This implies that it is not useful for the analyst to find ways to allow users to commit to a minimum precision if the practical situation at stake does not already include such a mechanism. 
In future work, we plan to analyze other decision-making structures, such as when agents make decisions asynchronously and can utilize information about the previous contribution levels by other agents.

In our basic model, we assume that the analyst can collect data from $n$ users at negligible cost. This assumption can be reasonable in scenarios where the data is already available in a repository, and the analyst merely has to inquire with individuals to contribute their data for secondary analysis. In this scenario, we showed that the population estimate's accuracy increases with $n$ (although each individual then lowers the precision of her contribution). We further extend the model to handle applications where there could be a more substantial cost of collecting data per user (e.g., cost of sending a survey). In that case, it is no longer optimal for the analyst to collect data from all users but we show, in the homogeneous case, how the analyst can then select the optimal number of users. The method outlined for the homogeneous case also provides a trajectory to approach the task of selecting the optimal set of agents to solicit data from in the heterogeneous case,  utilizing the ordering assumption of Theorem~\ref{nonhomtheorem}. 
% in the homogeneous case, and we provide a trajectory to approach this task for the heterogeneous case (i.e., by utilizing the ordering assumption of Theorem~\ref{nonhomtheorem}).  
Further, our results regarding the benefits of a minimum precision level apply also to costly data acquisition. In future work, we will consider non-linear cost (e.g., concave) to further generalize our results.

%Still, once the analyst has selected the optimal number of users to solicit data from (which we show how to do), all of our results readily apply to such a subpopulation and we show how to improve the estimation at no extra cost by restricting the agents' strategy set. In the case of heterogeneous agents, the analyst will need to solve the problem of selecting which agents to solicit data from based on their specific characteristics (or what is known about them by the analyst). This problem could be more complex (although it simplifies with the ordering assumption of Theorem~\ref{nonhomtheorem}) and is a potential option for future work. Similarly, we defer to future work the analysis of the same setting when the cost-per-agent is not linear (e.g., concave). However, these modifications would not alter our main results on the restriction of the agents' strategy space. 

A unique Nash equilibrium exists for all considered cases and extensions. Computing the exact equilibrium strategies may be non-trivial for agents in practice. However, knowledge about the uniqueness of the optimal strategies suggests the possibility of reaching the equilibrium via \textit{tacit coordination} when agents gain experience with comparable data contribution decisions \cite{Huyck90}. In addition, providing a minimum precision level will further guide agents in their decision-making.%However, this requires a further analysis of the model, that we let as future work.

In this paper, we consider the problem of estimating the population average of a single scalar quantity. However, the results also serve as building blocks to tackle more complex scenarios. For example, an analyst may need to estimate averages of several quantities which are not independent (if the quantities are independent, our results readily apply by considering several independent instances of the model, possibly with different privacy costs). Further, the analyst may want to estimate the parameter of a linear model as in \cite{ioannidis2013linear}. In both cases, the problem of selecting the users to solicit data from will become combinatorial and requires further study to find a suitable approximation. However, our techniques to characterize the equilibrium of the modified game will extend and will be instrumental in establishing the optimal strategy space to impose for a given set of users.

\section*{Acknowledgments}
This work was partially funded by the French Government (National Research Agency, ANR) through the ``Investments for the Future'' Program reference \#  ANR-11-LABX-0031-01; and by the Institut Mines-T\'el\'ecom through the ``Futur \& Rupture'' program. Jens Grossklags gratefully acknowledges the hospitality and support received as a Visiting Scientist at EURECOM during the earlier stages of this work. In addition, the authors would like to thank the anonymous reviewers for their detailed and helpful comments.


%\section*{Acknowledgment}
% NO ACKS IN SUBMISSION VERSION



%\bibliographystyle{IEEEtran}
%\bibliography{references,experimental_design_references}

%\begin{thebibliography}{1}
  %0.5em minus 0.4em\relax Harlow, England: Addison-Wesley, 1999.
%
%\end{thebibliography}

\bibliographystyle{IEEEtran}
\bibliography{references,experimental_design_references}

\begin{thebibliography}{10}
\providecommand{\url}[1]{#1}
\csname url@samestyle\endcsname
\providecommand{\newblock}{\relax}
\providecommand{\bibinfo}[2]{#2}
\providecommand{\BIBentrySTDinterwordspacing}{\spaceskip=0pt\relax}
\providecommand{\BIBentryALTinterwordstretchfactor}{4}
\providecommand{\BIBentryALTinterwordspacing}{\spaceskip=\fontdimen2\font plus
\BIBentryALTinterwordstretchfactor\fontdimen3\font minus
  \fontdimen4\font\relax}
\providecommand{\BIBforeignlanguage}[2]{{%
\expandafter\ifx\csname l@#1\endcsname\relax
\typeout{** WARNING: IEEEtran.bst: No hyphenation pattern has been}%
\typeout{** loaded for the language `#1'. Using the pattern for}%
\typeout{** the default language instead.}%
\else
\language=\csname l@#1\endcsname
\fi
#2}}
\providecommand{\BIBdecl}{\relax}
\BIBdecl

\bibitem{hmi2000}
P.~Lyman and H.~Varian, ``How much information?'' 2000, available at:
  \url{http://www2.sims.berkeley.edu/research/projects/how-much-info/}.

\bibitem{hmi2003}
------, ``How much information 2003?'' 2003, available at:
  \url{http://www2.sims.berkeley.edu/research/projects/how-much-info-2003/}.

\bibitem{Facebook2012}
J.~Constine, ``How big is {F}acebook's data? 2.5 billion pieces of content and
  500+ terabytes ingested every day,'' \emph{Techcrunch}, 2012.

\bibitem{WEC11}
{World Economic Forum, and Bain \& Company}, \emph{Personal Data: {T}he
  Emergence of a New Asset Class}.\hskip 1em plus 0.5em minus 0.4em\relax World
  Economic Forum, 2011.

\bibitem{Varian2014}
H.~Varian, ``Beyond big data,'' \emph{Business Economics}, vol.~49, no.~1, pp.
  27--31, 2014.

\bibitem{Solove06}
D.~Solove, ``A taxonomy of privacy,'' \emph{University of Pennsylvania Law
  Review}, vol. 154, no.~3, pp. 477--560, Jan. 2006.

\bibitem{Altman75}
I.~Altman, \emph{The Environment and Social Behavior}.\hskip 1em plus 0.5em
  minus 0.4em\relax Belmont, 1975.

\bibitem{Warren1890}
S.~Warren and L.~Brandeis, ``{The Right to Privacy},'' \emph{Harvard Law
  Review}, pp. 193--220, 1890.

\bibitem{Westin70}
A.~Westin, \emph{Privacy and freedom}.\hskip 1em plus 0.5em minus 0.4em\relax
  New York: Atheneum, 1970.

\bibitem{Acquisti13}
A.~Acquisti and C.~Fong, ``An experiment in hiring discrimination via online
  social networks,'' Carnegie Mellon University, Tech. Rep., 2013, available at
  SSRN: http://ssrn.com/abstract=2031979.

\bibitem{Mikians13}
J.~Mikians, L.~Gyarmati, V.~Erramilli, and N.~Laoutaris, ``Crowd-assisted
  search for price discrimination in e-commerce: {F}irst results,'' in
  \emph{Proceedings of the Conference on Emerging Networking Experiments and
  Technologies (CoNEXT)}, 2013, pp. 1--6.

\bibitem{Acquisti05}
A.~Acquisti and J.~Grossklags, ``Privacy and rationality in individual decision
  making,'' \emph{IEEE Security \& Privacy}, vol.~3, no.~1, pp. 26--33, 2005.

\bibitem{Kass03}
N.~Kass, M.~Natowicz, S.~Hull, R.~Faden, L.~Plantinga, L.~Gostin, and
  J.~Slutsman, ``The use of medical records in research: {W}hat do patients
  want?'' \emph{Journal of Law, Medicine \& Ethics}, vol.~31, pp. 429--433,
  2007.

\bibitem{Damschroder07}
L.~Damschroder, J.~Pritts, M.~Neblo, R.~Kalarickal, J.~Creswell, and
  R.~Hayward, ``Patients, privacy and trust: {P}atients' willingness to allow
  researchers to access their medical records,'' \emph{Social Science \&
  Medicine}, vol.~64, no.~1, pp. 223--235, 2007.

\bibitem{Robling04}
M.~Robling, K.~Hood, H.~Houston, R.~Pill, J.~Fay, and H.~Evans, ``Public
  attitudes towards the use of primary care patient record data in medical
  research without consent: {A} qualitative study,'' \emph{Journal of Medical
  Ethics}, vol.~30, no.~1, pp. 104--109, 2004.

\bibitem{Willison07}
D.~Willison, L.~Schwartz, J.~Abelson, C.~Charles, M.~Swinton, D.~Northrup, and
  L.~Thabane, ``Alternatives to project-specific consent for access to personal
  information for health research. {W}hat do canadians think?'' in
  \emph{Presentation at the 29th International Conference of Data Protection
  and Privacy Commissioners}, 2007.

\bibitem{Westin07}
A.~Westin, ``How the public views privacy and health research,'' 2007,
  {I}nstitute of {M}edicine.

\bibitem{ioannidis2013linear}
S.~Ioannidis and P.~Loiseau, ``Linear regression as a non-cooperative game,''
  in \emph{Web and Internet Economics}, ser. Lecture Notes in Computer Science,
  Y.~Chen and N.~Immorlica, Eds.\hskip 1em plus 0.5em minus 0.4em\relax
  Springer Berlin Heidelberg, 2013, vol. 8289, pp. 277--290.

\bibitem{Spiekermann01}
S.~Spiekermann, J.~Grossklags, and B.~Berendt, ``E-privacy in 2nd generation
  e-commerce: Privacy preferences versus actual behavior,'' in
  \emph{Proceedings of the 3rd ACM Conference on Electronic Commerce}, 2001,
  pp. 38--47.

\bibitem{Chessa15a}
M.~Chessa, J.~Grossklags, and P.~Loiseau, ``A short paper on incentives to
  share private information for population estimates,'' in \emph{Proceedings of
  the 19th International Conference on Financial Cryptography and Data Security
  (FC)}, 2015.

\bibitem{pukelsheim2006optimal}
F.~Pukelsheim, \emph{Optimal design of experiments}.\hskip 1em plus 0.5em minus
  0.4em\relax New York: Wiley, 1993.

\bibitem{atkinson2007optimum}
A.~Atkinson, A.~Donev, and R.~Tobias, \emph{Optimum experimental designs, with
  SAS}.\hskip 1em plus 0.5em minus 0.4em\relax Oxford University Press New
  York, 2007.

\bibitem{horel2013budget}
T.~Horel, S.~Ioannidis, and S.~Muthukrishnan, ``Budget feasible mechanisms for
  experimental design,'' in \emph{{LATIN 2014: Theoretical Informatics}}, ser.
  Lecture Notes in Computer Science, A.~Pardo and A.~Viola, Eds.\hskip 1em plus
  0.5em minus 0.4em\relax Springer Berlin Heidelberg, 2014, vol. 8392, pp.
  719--730.

\bibitem{RothSchoenebeck}
A.~Roth and G.~Schoenebeck, ``Conducting truthful surveys, cheaply,'' in
  \emph{Proceedings of the 13th ACM Conference on Electronic Commerce (EC)},
  2012, pp. 826--843.

\bibitem{Riederer}
C.~Riederer, V.~Erramilli, A.~Chaintreau, B.~Krishnamurthy, and P.~Rodriguez,
  ``For sale : Your data: By : You,'' in \emph{Proceedings of the 10th ACM
  Workshop on Hot Topics in Networks}, 2011, pp. 13:1--13:6.

\bibitem{Bilogrevic}
I.~Bilogrevic, J.~Freudiger, E.~De~Cristofaro, and E.~Uzun, ``What's the gist?
  {P}rivacy-preserving aggregation of user profiles,'' in \emph{Computer
  Security - ESORICS 2014}, ser. Lecture Notes in Computer Science,
  M.~Kutylowski and J.~Vaidya, Eds.\hskip 1em plus 0.5em minus 0.4em\relax
  Springer International Publishing, 2014, vol. 8713, pp. 128--145.

\bibitem{Dwork06}
C.~Dwork, ``Differential privacy,'' in \emph{Automata, Languages and
  Programming}, ser. Lecture Notes in Computer Science, M.~Bugliesi,
  B.~Preneel, V.~Sassone, and I.~Wegener, Eds.\hskip 1em plus 0.5em minus
  0.4em\relax Springer Berlin Heidelberg, 2006, vol. 4052, pp. 1--12.

\bibitem{kifer2012private}
D.~Kifer, A.~Smith, and A.~Thakurta, ``Private convex empirical risk
  minimization and high-dimensional regression,'' \emph{JMLR W{\&}CP
  (Proceedings of COLT 2012)}, vol.~23, pp. 25.1--25.40, 2012.

\bibitem{ghosh-roth:privacy-auction}
A.~Ghosh and A.~Roth, ``Selling privacy at auction,'' in \emph{Proceedings of
  the 12th ACM Conference on Electronic Commerce}, 2011, pp. 199--208.

\bibitem{approximatemechanismdesign}
K.~Nissim, R.~Smorodinsky, and M.~Tennenholtz, ``Approximately optimal
  mechanism design via differential privacy,'' in \emph{Proceedings of the 3rd
  Innovations in Theoretical Computer Science Conference}, 2012, pp. 203--213.

\bibitem{roth-liggett}
K.~Ligett and A.~Roth, ``{Take It or Leave It: Running a Survey When Privacy
  Comes at a Cost},'' in \emph{Internet and Network Economics}, ser. Lecture
  Notes in Computer Science, P.~Goldberg, Ed.\hskip 1em plus 0.5em minus
  0.4em\relax Springer Berlin Heidelberg, 2012, vol. 7695, pp. 378--391.

\bibitem{Chen13}
Y.~Chen, S.~Chong, I.~Kash, T.~Moran, and S.~Vadhan, ``Truthful mechanisms for
  agents that value privacy,'' in \emph{Proceedings of the Fourteenth ACM
  Conference on Electronic Commerce (EC)}, 2013, pp. 215--232.

\bibitem{vaidya2005privacy}
J.~Vaidya, C.~Clifton, and Y.~Zhu, \emph{Privacy Preserving Data Mining}.\hskip
  1em plus 0.5em minus 0.4em\relax Springer, 2006.

\bibitem{domingo2008survey}
J.~Domingo-Ferrer, ``A survey of inference control methods for
  privacy-preserving data mining,'' in \emph{Privacy-Preserving Data Mining},
  ser. Advances in Database Systems, C.~Aggarwal and P.~Yu, Eds.\hskip 1em plus
  0.5em minus 0.4em\relax Springer, 2008, vol.~34, pp. 53--80.

\bibitem{agrawal2000privacy}
R.~Agrawal and R.~Srikant, ``Privacy-preserving data mining,'' in
  \emph{Proceedings of the 2000 ACM SIGMOD International Conference on
  Management of Data}, 2000, pp. 439--450.

\bibitem{oliveira2003privacy}
S.~Oliveira and O.~Zaiane, ``Privacy preserving clustering by data
  transformation,'' in \emph{Proceedings of the XVIII Simposio Brasileiro de
  Bancos de Dados}, 2003, pp. 304--318.

\bibitem{atallah1999disclosure}
M.~Atallah, E.~Bertino, A.~Elmagarmid, M.~Ibrahim, and V.~Verykios,
  ``Disclosure limitation of sensitive rules,'' in \emph{Proceedings of the
  Workshop on Knowledge and Data Engineering Exchange (KDEX'99)}, 1999, pp.
  45--52.

\bibitem{Duchi13a}
J.~Duchi, M.~Jordan, and M.~Wainwright, ``Local privacy and statistical minimax
  rates,'' in \emph{Proceedings of the 54th IEEE Annual Symposium on
  Foundations of Computer Science (FOCS)}, 2013, pp. 429--438.

\bibitem{Cummings15}
R.~Cummings, K.~Ligett, A.~Roth, Z.~Wu, and J.~Ziani, ``Accuracy for sale:
  {A}ggregating data with a variance constraint,'' in \emph{Proceedings of the
  Conference on Innovations in Theoretical Computer Science (ITCS)}, 2015, pp.
  317--324.

\bibitem{Aperjis14}
C.~Aperjis, V.~Gkatzelis, and B.~Huberman, ``Pricing private data,''
  \emph{Electronic Markets}, {forthcoming}.

\bibitem{dekel2010incentive}
O.~Dekel, F.~Fischer, and A.~D. Procaccia, ``Incentive compatible regression
  learning,'' \emph{Journal of Computer and System Sciences}, vol.~76, no.~8,
  pp. 759--777, 2010.

\bibitem{perote2004}
J.~Perote and J.~Perote-Pena, ``Strategy-proof estimators for simple
  regression,'' \emph{Mathematical Social Sciences}, vol.~47, no.~2, pp.
  153--176, 2004.

\bibitem{Cai14a}
Y.~Cai, C.~Daskalakis, and C.~Papadimitriou, ``Optimum statistical estimation
  with strategic data sources,'' 2014, preprint, available as arXiv:1408.2539.

\bibitem{Morgan00a}
J.~Morgan, ``Financing public goods by means of lotteries,'' \emph{Review of
  Economic Studies}, vol.~67, no.~4, pp. 761--84, 2000.

\bibitem{Biczok13}
G.~Bicz{\'o}k and P.~Chia, ``Interdependent privacy: {L}et me share your
  data,'' in \emph{Financial Cryptography and Data Security}, ser. Lecture
  Notes in Computer Science, A.-R. Sadeghi, Ed.\hskip 1em plus 0.5em minus
  0.4em\relax Springer, 2013, vol. 7859, pp. 338--353.

\bibitem{pu2014economic}
Y.~Pu and J.~Grossklags, ``An economic model and simulation results of app
  adoption decisions on networks with interdependent privacy consequences,'' in
  \emph{Decision and Game Theory for Security}, R.~Poovendran and W.~Saad,
  Eds.\hskip 1em plus 0.5em minus 0.4em\relax Springer, 2014, vol. 8840, pp.
  246--265.

\bibitem{Backes12}
M.~Backes, A.~Kate, M.~Maffei, and K.~Pecina, ``Obliviad: {P}rovably secure and
  practical online behavioral advertising,'' in \emph{Proceedings of the IEEE
  Symposium on Security and Privacy}, 2012, pp. 257--271.

\bibitem{Guha11a}
S.~Guha, B.~Cheng, and P.~Francis, ``Privad: Practical privacy in online
  advertising,'' in \emph{Proceedings of the 8th USENIX Conference on Networked
  Systems Design and Implementation (NSDI)}, 2011, pp. 169--182.

\bibitem{Nardi12}
Y.~Nardi, S.~Fienberg, and R.~Hall, ``Achieving both valid and secure logistic
  regression analysis on aggregated data from different private sources,''
  \emph{Journal of Privacy and Confidentiality}, vol.~4, no.~1, 2012.

\bibitem{Hall12}
R.~Hall, S.~Fienberg, and Y.~Nardi, ``Secure multiple linear regression based
  on homomorphic encryption,'' \emph{Journal of Official Statistics}, vol.~27,
  no.~4, pp. 669--691, 2011.

\bibitem{Naor99a}
M.~Naor, B.~Pinkas, and R.~Sumner, ``Privacy preserving auctions and mechanism
  design,'' in \emph{Proceedings of the 1st ACM Conference on Electronic
  Commerce (EC)}, 1999, pp. 129--139.

\bibitem{Izmalkov05a}
S.~Izmalkov, S.~Micali, and M.~Lepinski, ``Rational secure computation and
  ideal mechanism design,'' in \emph{Proceedings of the 46th IEEE Annual
  Symposium on Foundations of Computer Science (FOCS)}, 2005, pp. 585--594.

\bibitem{Cranor02}
L.~Cranor, \emph{Web privacy with {P3P} - {T}he platform for privacy
  preferences}.\hskip 1em plus 0.5em minus 0.4em\relax O'Reilly, 2002.

\bibitem{Berthold01}
O.~Berthold and M.~K\"{o}hntopp, ``Identity management based on {P3P},'' in
  \emph{Designing Privacy Enhancing Technologies}, ser. Lecture Notes in
  Computer Science, H.~Federrath, Ed.\hskip 1em plus 0.5em minus 0.4em\relax
  Springer, Berlin Heidelberg, 2001, vol. 2009, pp. 141--160.

\bibitem{Cranor02b}
L.~Cranor, M.~Arjula, and P.~Guduru, ``Use of a {P3P} user agent by early
  adopters,'' in \emph{Proceedings of the ACM Workshop on Privacy in the
  Electronic Society}, 2002, pp. 1--10.

\bibitem{Wang07}
Y.~Wang and A.~Kobsa, ``Respecting users' individual privacy constraints in web
  personalization,'' in \emph{User Modeling 2007}, ser. Lecture Notes in
  Computer Science, C.~Conati, K.~McCoy, and G.~Paliouras, Eds.\hskip 1em plus
  0.5em minus 0.4em\relax Springer Berlin Heidelberg, 2007, vol. 4511, pp.
  157--166.

\bibitem{Acquisti12}
A.~Acquisti and J.~Grossklags, ``An online survey experiment on ambiguity and
  privacy,'' \emph{Communications \& Strategies}, vol.~49, no.~4, pp. 19--39,
  2012.

\bibitem{Acquisti13b}
A.~Acquisti, L.~John, and G.~Loewenstein, ``What is privacy worth?''
  \emph{Journal of Legal Studies}, vol.~42, no.~2, 2013.

\bibitem{Grossklags07}
J.~Grossklags and A.~Acquisti, ``When 25 cents is too much: {A}n experiment on
  willingness-to-sell and willingness-to-protect personal information,'' in
  \emph{Proceedings of the Workshop on the Economics of Information Security},
  2007.

\bibitem{Acquisti13c}
A.~Acquisti, I.~Adjerid, and L.~Brandimarte, ``Gone in 15 seconds: {T}he limits
  of privacy transparency and control,'' \emph{IEEE Security \& Privacy},
  vol.~11, no.~4, pp. 72--74, 2013.

\bibitem{Brandimarte12}
L.~Brandimarte, A.~Acquisti, and G.~Loewenstein, ``Misplaced confidences:
  {P}rivacy and the control paradox,'' \emph{Social Psychological and
  Personality Science}, vol.~4, no.~3, pp. 340--347, 2013.

\bibitem{Wang11}
N.~Wang, H.~Xu, and J.~Grossklags, ``Third-party apps on {F}acebook: {P}rivacy
  and the illusion of control,'' in \emph{Proceedings of the 5th ACM Symposium
  on Computer Human Interaction for Management of Information Technology},
  2011.

\bibitem{Wang13}
N.~Wang, J.~Grossklags, and H.~Xu, ``An online experiment of privacy
  authorization dialogues for social applications,'' in \emph{Proceedings of
  the 2013 Conference on Computer Supported Cooperative Work (CSCW)}, 2013, pp.
  261--272.

\bibitem{Huberman05}
B.~Huberman, E.~Adar, and L.~Fine, ``Valuating privacy,'' \emph{IEEE Security
  \& Privacy}, vol.~3, no.~5, pp. 22--25, 2005.

\bibitem{Spiekermann12}
C.~Bauer, J.~Korunovska, and S.~Spiekermann, ``On the value of information:
  What {F}acebook users are willing to pay,'' in \emph{Proceedings of the 33rd
  International Conference on Information Systems}, 2012.

\bibitem{Huyck90}
J.~Van~Huyck, R.~Battalio, and R.~Beil, ``Tacit coordination games, strategic
  uncertainty, and coordination failure,'' \emph{The American Economic Review},
  vol.~80, no.~1, pp. 234--248, 1990.

\bibitem{Milgrom94a}
P.~Milgrom and J.~Roberts, ``Comparing equilibria,'' \emph{American Economic
  Review}, vol.~84, pp. 441--459, 1994.

\end{thebibliography}

%\balance
\appendix
\label{appendix}
\section{Proofs}

\subsection{Corollary 1 from ``Comparing Equilibria'' of Milgrom and Roberts \cite{Milgrom94a}}\label{milgrom}

Many of our theoretical contributions rely on a result of the paper from Milgrom and Roberts, ``Comparing Equilibria''. To help the reader, we present here this result. For simplicity, we replace the hypothesis of ``continuous but for upward jumps'', with the stronger hypothesis of ``continuous'', which is verified by our functions to which we apply the theorem. We also adapt the statement of the corollary to a fixed point problem defined on a generic interval $[a,b]\subset \mathbb R+$.

\begin{corollary}[Milgrom and Roberts]
Let $g(x,t): [a,b]\times T \rightarrow [a,b]$, where $T$ is any partially ordered set. Suppose that for all $t\in T$, $g$ is continuous in $x$. Then $x_L(t) = \textrm{inf} \{x|g(x,t)\le x\}$ and $x_H(t) = \textrm{sup} \{x|g(x,t)\ge x\}$ are the extreme fixed points of $g$, that is, the lowest and the highest solutions of the equation $g(x,t)=x$. If, in addition, $g$ is monotone non-decreasing in $t$ for all $x\in [0,1]$, then the functions $x_L(\cdot)$ and $x_H(\cdot)$ are monotone non-decreasing, and if $g$ is strictly increasing in $t$, then these functions are strictly increasing.
\end{corollary}

\subsection{Proof of Theorem~\ref{theorem1stage}}\label{theorem1stageproof}

$\Gamma$ is a symmetric potential game, with potential function $\Phi:[0,1/\sigma^2]^n \rightarrow \bar{\mathbb R}$, s.t., for each $\lambdab\in [0,1/\sigma^2]^n$
\begin{align}\label{potentialfunction1stage}
\Phi(\lambdab) = \sum_{j\in N}c(\lambda_j) + f(\lambdab).
\end{align}
By the definition of potential game, the set of Nash equilibria of $\Gamma$ is contained in the set of local minima of function $\Phi$. Then, as function $\Phi$ has a unique local minimum $\lambdab^*\in [0,1/\sigma^2]^n$, it has to coincide with the unique Nash equilibrium of $\Gamma$. In particular, the optimum $\lambdab^*$ is such that $\lambdab^*\neq (0,\ldots ,0)$ and for each $i\in N$, $\lambda_i^*$ satisfies the following KKT conditions
\begin{align}\label{KKTconditions}
\left\{
\begin{aligned}
& -\frac{1}{(\sum_{j\in N}\lambda_j^*)^2}F'\left( \frac{1}{\sum_{j\in N}\lambda_j^*} \right)+c'(\lambda_i^*)-\psi_i^*+\phi_i^* = 0 \\
& \psi_i^* \lambda_i^*=0\	\	\phi_i^*(\lambda_i^*-1/\sigma^2)=0,\	\	 \psi_i^*,\phi_i^*\ge 0.
\end{aligned}
\right.
\end{align}
Observe that, as a consequence of the assumption that $c'(0)=0$, $\lambda_i^*>0$ for each $i\in N$. In fact, if we suppose there exists $i\in N$ s.t. $\lambda_i^*=0$, the $i$th-equation of the KKT conditions cannot be satisfied, as
$$
-\frac{1}{(\sum_{j\in N}\lambda_j^*)^2}F'\left( \frac{1}{\sum_{j\in N}\lambda_j^*} \right)-\psi_i^* < 0,
$$
because $\psi_i^*>0$ and $F'>0$ as $F$ is strictly convex. Moreover, as $\Phi$ is a symmetric function on a symmetric domain, the only minimum is symmetric, i.e., $\lambda_i^*=\lambda^*$ for each $i\in N$.

\subsection{Proof of Proposition~\ref{DecPrec}}\label{DecPrecproof}

From \eqref{KKTconditions}, $\lambda^*$ is the unique solution of the following fixed point problem
\begin{align}\label{fixedpoint1}
\lambda=g(n,\lambda),
\end{align}
where function $g: \mathbb N_+ \times [0,1/\sigma^2] \rightarrow [0,+\infty]$ is defined for each $\lambda\in (0,1/\sigma^2]$ and for each $n\in \mathbb N_+$ as
\begin{align}\label{fixedpoint2}
g(n,\lambda) = \min{\left\{\sqrt{F'\left( \frac{1}{n\lambda} \right) \frac{1}{n^2c'(\lambda)}} , 1/\sigma^2 \right\}}
\end{align}
and by continuity as $\lim_{\lambda \rightarrow 0^+}g(n,\lambda)$ in $\lambda=0$ for each $n\in \mathbb N_+$.

We consider this fixed point problem, but with the parameter $n$ defined on the real interval $[1,+\infty]$. For each $n\in [1,+\infty]$, $g$ is continuous in $\lambda$. Function $g$ is monotonic non-increasing in $n$. In fact,
\begin{align}
\nonumber \frac{\partial{g}}{\partial{n}}= & \frac{1}{2\sqrt{F'\left( \frac{1}{n\lambda} \right) \frac{1}{n^2c'(\lambda)}}} \\
\nonumber & \left[-\frac{1}{n^4\lambda^2 c'(\lambda)}F''\left(\frac{1}{n \lambda}\right)-\frac{2nc'(\lambda)}{n^4c'(\lambda)^2}F'\left(\frac{1}{n\lambda}\right)\right] \\
\nonumber & < 0
\end{align}
Applying Corollary 1 of Milgrom and Roberts \cite{Milgrom94a}, recalled in Appendix~\ref{milgrom}, (with parameter $t=-n$) the unique fixed point $\lambda^*(n)$ is non-increasing in $n$, and this proves \textit{(i)}.

To prove \textit{(ii)}, we observe that
$$
\lim_{n\rightarrow +\infty}g(n,\lambda)=0 \textrm{ (zero function)},
$$
and the unique fixed point of the zero function is $0$.

\subsection{Proof of Proposition~\ref{DecEstCost}}\label{DecEstCostproof}

If $\lambda^*(n) = 1/\sigma^2$ or $\lambda^*(n+1) = 1/\sigma^2$, \textit{(i)} is trivial. Suppose that both $\lambda^*(n)\neq 1/\sigma^2$ and $\lambda^*(n+1)\neq 1/\sigma^2$. Moreover, suppose by contradiction that 
\begin{align}\label{eq1}
\frac{1}{n\lambda^*(n)}<\frac{1}{(n+1)\lambda^*(n+1)}.
\end{align}
It follows that
\begin{align}\label{eq2}
F'\left(\frac{1}{n\lambda^*(n)}\right) < F'\left(\frac{1}{(n+1)\lambda^*(n+1)}\right)
\end{align}
because of the strict convexity of $F$. Moreover, as $\lambda^*(n)>\lambda^*(n+1)$ (by Corollary~\ref{DecPrec}), it follows that
\begin{align}\label{eq3}
c'(\lambda^*(n)) > c'(\lambda^*(n+1))
\end{align}
because of the strict convexity of $c$. From \eqref{eq2} and \eqref{eq3}, it follows that
\begin{align*}
& \frac{1}{n\lambda^*(n)} \\
& = \frac{1}{n\sqrt{F'\left(\frac{1}{n\lambda^*(n)}\right)\frac{1}{n^2c'(\lambda^*(n))}}} \\
& > \frac{1}{(n+1)\sqrt{F'\left(\frac{1}{(n+1)\lambda^*(n+1)}\right)\frac{1}{(n+1)^2c'(\lambda^*(n+1))}}} \\
& = \frac{1}{(n+1)\lambda^*(n+1)},
\end{align*}
which contradicts $\eqref{eq1}$ and then proves \textit{(i)}. 

To prove \textit{(ii)}, observe that, because of Proposition \ref{DecPrec}-\textit{(ii)} and because of Assumption~\ref{privacyassumption},
\begin{align*}
\lim_{n\rightarrow +\infty}c'(\lambda^*(n))=0.
\end{align*}
If, by contradiction,
\begin{align}\label{eq5}
\lim_{n\rightarrow +\infty}\frac{1}{n\lambda^*(n)}>0,
\end{align}
then
\begin{align*}
\lim_{n\rightarrow +\infty}F'\left(\frac{1}{n\lambda^*(n)}\right)>0,
\end{align*}
because of the strict convexity of $F$, and consequently
\begin{align*}
\lim_{n\rightarrow +\infty}\frac{1}{n\sqrt{F'\left(\frac{1}{n\lambda^*(n)}\right)\frac{1}{n^2c'(\lambda^*(n))}}}=0,
\end{align*}
which contradicts \eqref{eq5} and then proves \textit{(ii)}

\subsection{Proof of Theorem~\ref{theoremeta}}\label{theoremetaproof}

First, observe that for each $S\subseteq N$, with $s\ge 1$, and for each $\eta\in [0,1/\sigma^2]$, the game $\Gamma(S,\eta)$ is still a potential game, with potential function $\Phi$ as in \eqref{potentialfunction1stage}, but defined on the domain $\big[\{0\}\cup [\eta,1/\sigma^2]\big]^s$. The set of Nash equilibria of $\Gamma(S,\eta)$ is included in the set of the local minima of $\Phi$ on this new domain.

When $s=1$, the potential function and the cost function of the only agent coincide. Then, a strategy profile is a Nash equilibrium if and only if it is a global minimum of $\Phi$.  If $\eta \le \lambda^*(1)$, then the only global minimum of $\Phi$ is still $\lambda^*(1)$. If $\eta > \lambda^*(1)$, then the only global minimum is $\eta$.

Now, let $s>1$. We define the function $\tilde{g}:\mathbb N_+ \times [0,1/\sigma^2] \rightarrow [0,+\infty]$ s.t., for each $\eta\in (0,1/\sigma^2]$ and for each $n\in \mathbb N_+$
\begin{align}\label{fixedpoint2eta}
\tilde{g}(s,\eta) = \min{\left\{\frac{F\left( \frac{1}{(s-1)\eta}\right)-F\left( \frac{1}{s\eta}\right)}{c(\eta)}\cdot \eta , 1/\sigma^2 \right\}}
\end{align}
and we extend it by continuity in $\eta=0$ for each $n\in \mathbb N_+$. We consider the following fixed point problem
\begin{align}\label{fixedpoint1eta}
\eta=\tilde{g}(s,\eta),
\end{align}

and we show that this fixed point problem has a unique solution. To prove that, we show first that equation
\begin{align}\label{fixedpoint3eta}
F\left( \frac{1}{(s-1)\eta}\right)-F\left( \frac{1}{s\eta}\right)=c(\eta)
\end{align}
in the $\eta$ variable, has at most one solution in $[0,1/\sigma^2]$. This can be seen by noticing that, for each $s>1$, the difference $\frac{1}{(s-1)\eta} - \frac{1}{s\eta}$ is decreasing in $\eta$. Moreover, function $F$ is strictly convex and non-increasing in $\eta$, and this implies that the difference 
\begin{align}\label{difference}
F\left( \frac{1}{(s-1)\eta}\right)-F\left( \frac{1}{s\eta}\right)
\end{align}
is a decreasing function of $\eta$. As $c$ is a non-decreasing function of $\eta$, it follows that \eqref{fixedpoint3eta} has at most one solution in the given interval.

The fixed point of \eqref{fixedpoint1eta} is given by this solution (if it exists), or by $1/\sigma^2$ otherwise, and then it is unique. We denote this unique fixed point by $\eta^*(s)$.

Looking for the Nash equilibria when $s>1$, at first we focus on the ones which are in $[\eta,1/\sigma^2]^s$. In particular, we can distinguish the three following subcases (observe that, in case we have that $\lambda^*(s) > \eta^*(s)$, this simply implies that the subcase \textit{(iib)} will never occur):

\begin{itemize}
\item[\textit{(ia)}] When $\eta\in [0,\lambda^*(s)]$, as $\lambdab^*(s)$ is the unique local minimum of the potential function on the domain $[0,1/\sigma^2]^s$ and as $\lambdab^*(s)\in [\eta,1/\sigma^2]^s$, then, because of the convexity, it is still the only local minimum of the potential function on $[\eta,1/\sigma^2]^s$. In particular, it is a Nash equilibrium of $\Gamma(S,\eta)$. In fact, if there exists a deviation of agent $i\in S$ for the game $\Gamma(S,\eta)$ which makes her cost function smaller, it would be a feasible deviation which makes her function smaller also for the game $\Gamma(S,0)$, and this would contradict the fact that $\lambdab^*(s)$ is a Nash equilibrium of $\Gamma(S,0)$.
\item[\textit{(ib)}] When $\eta\in (\lambda^*(s),\eta^*(s)]$, the vector $\etab=(\eta)_{i\in S}$ is the only local minimum of the potential function on $[\eta,1/\sigma^2]^s$. In particular, it is a Nash equilibrium. In fact, because of the strictly convexity of the potential function, any deviation of agent $i\in N$ to a precision level in $(\eta,1/\sigma^2]$ would make her cost function bigger. Moreover, if agent $i\in N$ deviates to $0$, her cost function cannot become smaller. In fact, we have that, from \eqref{fixedpoint3eta}, when $\eta \le \eta^*(s)$,
\begin{align*}
F\left( \frac{1}{(s-1)\eta}\right) \ge F\left( \frac{1}{s\eta}\right) + c(\eta).
\end{align*}
The term on the left represents the cost of agent $i$ deviating to zero, and the term on the right denotes the cost when she decides to keep on choosing a precision equal to $\eta$.
\item[\textit{(ii)}] When $\eta\in (\eta^*(s),1/\sigma^2]$, the only local minimum in $[\eta,1/\sigma^2]^s$ is $\etab$. But this is not a Nash equilibrium. In fact, still because of \eqref{fixedpoint3eta}, when $\eta > \eta^*(s)$
\begin{align*}
F\left( \frac{1}{(s-1)\eta}\right) < F\left( \frac{1}{s\eta}\right) + c(\eta),
\end{align*}
and this means that an agent can make her cost function smaller deviating to zero.
\end{itemize}
We can now remark that, as $\lambda^*(s)$ is a Nash equilibrium for $\Gamma(S,0)$, it implies that, by playing that strategy, the agents do not have incentives to deviate to zero. As $\eta^*(s)$ is the maximum minimum precision level s.t., if the agents are playing $\Gamma(S,\eta^*(s))$, they do not have incentives to deviate by $\eta^*(s)$, it follows that $\lambda^*(s)\le \eta^*(s)$.

We proved that when $\eta\in (\eta^*(s),1/\sigma^2]$, there does not exist a Nash equilibrium of $\Gamma(S,\eta)$ in $[\eta,1/\sigma^2]^s$, and this proves Theorem~\ref{theoremeta}-\textit{(ii)}. We have also proved that when $\eta\in [0,\eta^*(s)]$, there exists a unique Nash equilibrium of $\Gamma(S,\eta)$ in $[\eta,1/\sigma^2]^s$. In order to prove that there do not exist other equilibria with a zero component (and then, in order to prove Theorem~\ref{theoremeta}-\textit{(i)}), we first state the following lemma.

\begin{lemma}\label{lemmaapp1}
Suppose that $\lambdab'=(\lambda'_1,\ldots ,\lambda'_s)$ is a local minimum of the potential function $\Phi$ on $\big[\{0\}\cup [\eta,1/\sigma^2]\big]^s$, with $\eta\in [0,1/\sigma^2]$ and call $T=\{i\in S: \lambda'_i=0\}$, with $t=|T|$. Then, $\lambdab'$ is a local minimum on $\{0\}^t\times [\eta,1/\sigma^2]^{s-t}$ and it is s.t. $\lambda'_i=\lambda'$ for each $i\in S\setminus T$, with
\begin{align}\label{eqapp1}
\lambda'=\left\{
\begin{aligned}
& \lambda^*(s-t) & \textrm{ if } 0 \le \eta \le \lambda^*(s-t) \\
& \eta & \textrm{ if } \lambda^*(s-t) < \eta \le 1/\sigma^2. \\
\end{aligned}
\right.
\end{align}
\end{lemma}

Suppose now that there exists a local minimum $\lambdab'$ s.t. $\lambda'_i=0$ for at least one $i\in S$ and call $T=\{i\in S: \lambda'_i=0\}$, with $t=|T|\ge 1$, the set of agents who are at a zero precision level. Then, because of Lemma \ref{lemmaapp1}, $\lambda'_i=\lambda'$ for each $i\in S\setminus T$ and it is given by \eqref{eqapp1}. We show that this cannot be a Nash equilibrium. In fact, we have that,
\begin{align}\label{deviation}
& c(\lambda') \le F\left( \frac{1}{(s-1)\lambda'}\right) - F\left( \frac{1}{s\lambda'}\right)\\
\nonumber & < F\left( \frac{1}{(s-t)\lambda'}\right) - F\left( \frac{1}{(s-t+1)\lambda'}\right),
\end{align}
when $t\ge 1$. The first inequality follows from \eqref{fixedpoint3eta} and from the fact that $\lambda' \in [0,\eta^*(s)]$. The second inequality follows from the fact that, fixed $\eta$, the difference in \eqref{difference} is a decreasing function also of $s$. From \eqref{deviation}, it follows that if an agent in $S\setminus T$ deviates moving from the precision level $\eta$ to the precision level $\lambda'$, she can strictly decrease her cost function.

This proves that any local minimum of $\Phi$ s.t. at least one agent chooses a zero precision level, cannot be a Nash equilibrium. Then, when $s>1$, the equilibrium is unique and it is given by \eqref{solution1stageeta2}, with $\eta^*(s)$ unique solution of \eqref{fixedpoint1eta}.
%\begin{align*}
%\frac{1}{(n-t)\lambda'} + c\lambda'^k > \frac{1}{(n-t-1)\lambda'} 
%\end{align*}
%for each $\eta \in (\bar{\eta}(n-t),1/\sigma^2]$, when $t\ge 1$, meaning that an agent in $N\setminus T$ has incentives to deviate to $0$.

\subsection{Proof of Theorem~\ref{opteta1stage}}\label{opteta1stageproof}

We have already seen in the proof of Theorem~\ref{theoremeta}, that for each $S\subseteq N$, $\lambda^*(s) \le \eta^*(s)$. It follows that $\sigma_M^2(\etab^*(s)) \le \sigma_M^2(\lambdab^*(s))$. This means that, fixed the number of agents $s$, it is optimal, for the analyst, to choose a minimum precision level equal to $\eta^*(s)$.

\textbf{Step 1:} First, we show now that, if $\lambda^*(s)\neq 1/\sigma^2$, this inequality is strict, meaning that $\lambda^*(s) < \eta^*(s)$ and the analyst can strictly improve the estimation, by choosing $\eta^*(s)$ instead of $0$ as minimum precision level. In fact, if  $\lambda^*(s) = \eta^*(s)$, it follows that $\lambda^*(s)$ is s.t.
$$
c(\lambda^*(s)) = F\left( \frac{1}{(s-1)\lambda^*(s)}\right) - F\left(\frac{1}{s\lambda^*(s)}\right).
$$
But then, at equilibrium, the potential function $\Phi$ is s.t.
\begin{align*}
& F\left( \frac{1}{s\lambda^*(s)}\right) + sc(\lambda^*(s)) \\
& = F\left( \frac{1}{s\lambda^*(s)}\right) + (s-1)c(\lambda^*(s)) \\
& + F\left( \frac{1}{(s-1)\lambda^*(s)}\right) - F\left(\frac{1}{s\lambda^*(s)}\right) \\
& = F\left( \frac{1}{(s-1)\lambda^*(s)}\right) + (s-1)c(\lambda^*(s)).
\end{align*}
This implies that the potential function is minimal for an agent $i$ choosing $\lambda_i^*(S) = 0$, and this contradicts the fact that the equilibrium of $\Phi$ is unique, symmetric and s.t. $\lambdab^*\neq (0,\ldots , 0)$. It follows that, for each $S\subseteq N$, $\lambda^*(s) < \eta^*(s)$.

\textbf{Step 2:} Second, we observe that $\eta^*(s)$ is nonincreasing in $s$. This because $\eta^*(s)$ is the unique fixed point of the problem in \eqref{fixedpoint1eta}, and the function $\tilde{g}(s,\eta)$ is continuous, nondecreasing in $\eta$ and nonincreasing in $s$. Then, applying Corollary 1 of Milgrom and Roberts \cite{Milgrom94a}, recalled in Appendix~\ref{milgrom}, to it (with parameter $t=-s$), we have the result.
%to the problem and to the function defined in \eqref{fixedpoint2eta} satisfies the hypothesis of
%$$
%\lim_{n\rightarrow +\infty}\tilde{g}(s,\eta)=0 \textrm{ (zero function)},
%$$
%then the unique fixed point of the zero function is $0$.

%\textbf{Step 3:} (Pay attention that this step is not used here after, but it could be useful for other properties...) Third, we observe that
%$$
%\lim_{n\rightarrow +\infty}\sigma^2_M(\etab^*(s))=0.
%$$
%In fact, as $\lambda^*(s)<\eta^*(s)$, it follows that
%$$
%0\le \lim_{n\rightarrow +\infty} \sigma^2_M(\etab^*(s)) \le \lim_{n\rightarrow +\infty} \sigma^2_M(\lambdab^*(s)) \rightarrow_{n\rightarrow +\infty}0.
%$$

\textbf{Step 3:} Finally, we show that $\sigma^2_M(\etab^*(s))$ is nonincreasing in $s$, and then, that it is optimal, for the analyst, to collect data from the largest possible number of agents. We suppose, by contradiction, that there exists $k\in \mathbb N_+$ s.t. $\sigma^2_M(\etab^*(k))<\sigma^2_M(\etab^*(k+1))$, or equivalently s.t. $k\cdot \eta^*(k)>(k+1)\cdot \eta^*(k+1)$. We have shown in step 2 that $\eta^*(s)$ is nonincreasing in $s$, then $\eta^*(k)\ge \eta^*(k+1)$. Suppose $\eta^*(k)\neq 1/\sigma^2$ (otherwise, the result is trivial). By definition, $\eta^*(k)$ is the solution of \eqref{fixedpoint1eta} for $s=k$ and $\eta^*(k+1)$ for $s=k+1$. We write the equation in \eqref{fixedpoint1eta} as 
\begin{equation}\label{last}
F\left(\frac{1}{t_1-\eta^*(k)}\right)-F\left(\frac{1}{t_1}\right)=c(\eta^*(k))
\end{equation}
where $t_1=k\cdot \eta^*(k)$. Similarly, we write
$$
F\left(\frac{1}{t_0-\eta^*(k+1)}\right)-F\left(\frac{1}{t_0}\right)=c(\eta^*(k+1))
$$
where $t_0=(k+1)\cdot \eta^*(k+1)$. Because of the hypothesis by contradiction, $t_0<t_1$; moreover the difference on the left in \eqref{last} is increasing as a function of the parameter. We may apply a straightforward adaptation of Milgrom and Roberts' Corollary 1 \cite{Milgrom94a}, recalled in Appendix~\ref{milgrom}, (on the right we do not have a linear function of $\eta$ as in the original statement, but a strictly increasing function of $\eta$) and we obtain that $\eta^*(k)<\eta^*(k+1)$, contradicting what we have shown in Step 2.\\

We have shown that for the analyst it is not optimal to implement the game with only a subset of the agents. Moreover, for the analyst it is not optimal to choose a minimum precision level $\eta > \eta^*(n)$. In fact, in that case, as we have seen in Section~\ref{onestageetasec}, if there exists an equilibrium, it is an equilibrium s.t. only a subset of agents choose a non-zero precision level, and this leads back to the previous case.

\subsection{Proof of Theorem~\ref{nonhomtheorem}}\label{nonhomtheoremproof}
The proof follows the proof of Theorem~\ref{theorem1stage}. In particular, the unique Nash equilibrium satisfies the KKT conditions in \eqref{KKTconditions} (with heterogenous privacy costs), from which it still follows that, because of the assumption that $c_i'(0)=0$ for each $i\in N$, $\lambda_i^*\neq 0$ for each $i\in N$. Given $i\in N$, the corresponding equilibrium precision level is s.t.
\begin{align}\label{KKT2}
c_i'(\lambda_i^*) = \frac{F'(\sigma_M^2(\lambdab^*))}{(\sum_{j\in N}\lambda_j^*)^2},
\end{align}
if the solution is smaller than or equal to $1/\sigma^2$, or by $1/\sigma^2$ otherwise.

As the right term is the same for each $i\in N$, it immediately follows that, if the $c_i$'s are s.t. $c_1'(\lambda) \le ... \le c_n'(\lambda)$, for each $\lambda\in [0,1/\sigma^2]$, then $\lambda^*_n \le \ldots \le \lambda^*_1$.

\subsection{Proof of Proposition~\ref{DecPrecHet}}\label{DecPrecHetproof}
From Equation \eqref{KKT2}, as soon as agent $n+1$ enters the game, fixing the strategies of the other agents, the term on the right decreases. In order to balance the equality at best response, and because of the convexity of the privacy cost $c_i$, fixing the strategy of the other agents, each agent $i\in N$ will choose a precision level which is smaller then the precision level at best response, without agent $n+1$. As a consequence, at equilibrium, $\lambda_i^*(N\cup\{n+1\}) \le \lambda_i^*(N)$ for each $i\in N$.

\subsection{Proof of Proposition~\ref{DecEstCostHet}}\label{DecEstCostHetproof}
We write Equation \eqref{KKT2} as
\begin{align*}
\displaystyle \frac{c_i'(\lambda_i^*)}{F'(\sigma_M^2(\lambdab^*))} = \sigma_M^2(\lambdab^*).
\end{align*}
We suppose by contradiction that $\sigma_M^2(\lambdab^*(N\cup\{n+1\})) > \sigma_M^2(\lambdab^*(N))$. Then, $F'(\sigma_M^2(\lambdab^*(N\cup\{n+1\}))) > F'(\sigma_M^2(\lambdab^*(N)))$, because of the convexity of $F$. Moreover, from Proposition~\ref{DecPrecHet}, we know that $\lambda_i^*(N\cup\{n+1\}) \le \lambda_i^*(N)$, and then $c'(\lambda_i^*(N\cup\{n+1\})) \le c'(\lambda_i^*(N))$ because of the convexity of the $c_i$s. It follows that
\begin{align*}
& \sigma_M^2(\lambdab^*(N\cup\{n+1\})) \\
& = \displaystyle \frac{c_i'(\lambda_i^*(N\cup\{n+1\})}{F'(\sigma_M^2(\lambdab^*(N\cup\{n+1\})))} \\
& < \displaystyle \frac{c_i'(\lambda_i^*(N)}{F'(\sigma_M^2(\lambdab^*(N)))} \\
& = \displaystyle \sigma_M^2(\lambdab^*(N)),
\end{align*}
and this contradicts the supposition by contradiction.

\subsection{Proof of Theorem~\ref{theoremetaHet}}\label{theoremetaHetproof}

At first, we recall that we denote by $\lambdab^*(S)$ the unique Nash equilibrium of the game $\Gamma(S,0)$. Then, for each $S\subseteq N$, with $s\ge 1$, and for each $\eta\in [0,1/\sigma^2]$, we observe that the game $\Gamma(S,\eta)$ is still a potential game, with potential function $\Phi$ as in \eqref{potentialfunction1stage}, but defined on the domain $\big[\{0\}\cup [\eta,1/\sigma^2]\big]^s$. The set of Nash equilibria of $\Gamma(S,\eta)$ is included in the set of the local minima of $\Phi$ on this new domain.

When $s=1$, the potential function and the cost function of the only agent coincide. Then, a strategy profile is a Nash equilibrium if and only if it is a global minimum of $\Phi$.  If $\eta \le \lambda^*(\{1\})$, then the only global minimum of $\Phi$ is still $\lambda^*(\{1\})$. If $\eta > \lambda^*(\{1\})$, then the only global minimum is $\eta$.

Now, let $s>1$. By definition of Nash equilibrium, the unique NE $\lambdab^*(S)$ is s.t.
$$
c_n(\lambda_n^*(S)) \le F\left( \frac{1}{\sum_{j\in N,j\neq n}\lambda_j^*(S)}\right) - F\left( \frac{1}{\sum_{j\in N}\lambda_j^*(S)}\right),
$$
which translates the fact that agent $s$ does not have incentives to deviate to zero.

\textbf{Step 1:} First, we show that
$$
c_n(\lambda_n^*(S)) < F\left( \frac{1}{\sum_{j\in N,j\neq n}\lambda_j^*(S)}\right) - F\left( \frac{1}{\sum_{j\in N}\lambda_j^*(S)}\right).
$$
By contradiction, if
$$
c_n(\lambda_n^*(S)) = F\left( \frac{1}{\sum_{j\in N,j\neq n}\lambda_j^*(S)}\right) - F\left( \frac{1}{\sum_{j\in N}\lambda_j^*(S)}\right),
$$
then at equilibrium the potential function $\Phi$ is s.t.
\begin{align*}
& F\left( \frac{1}{\sum_{j\in N}\lambda_j^*(S)}\right) + \sum_{j\in N}c_j(\lambda_j^*(S)) \\
& = F\left( \frac{1}{\sum_{j\in N}\lambda_j^*(S)}\right) + \sum_{j\in N, j\neq n}c_j(\lambda_j^*(S)) \\
& + F\left( \frac{1}{\sum_{j\in N,j\neq n}\lambda_j^*(S)}\right) - F\left( \frac{1}{\sum_{j\in N}\lambda_j^*(S)}\right) \\
& = F\left( \frac{1}{\sum_{j\in N,j\neq n}\lambda_j^*(S)}\right) + \sum_{j\in N, j\neq n}c_j(\lambda_j^*(S)).
\end{align*}
This implies that the potential function is minimal for $\lambda_n^*(S) = 0$, and this contradicts the fact that the equilibrium is unique and s.t. no agent is playing zero.

\textbf{Step 2:} Now, let $\eta^*(S)$ be s.t.
\begin{align}\label{eq9}
& c_n(\lambda_n^*(S,\eta^*(S)))\\
\nonumber & = F\left( \frac{1}{\sum_{j\in N,j\neq n}\lambda_j^*(S,\eta^*(S))}\right) - F\left( \frac{1}{\sum_{j\in N}\lambda_j^*(S,\eta^*(S))}\right),
\end{align}
where $\lambdab^*(S,\eta^*(S))$ is the local minimum of $\Phi$ on $[\eta^*(S), 1/\sigma^2]^s$. Note that this $\eta^*(S)$ is unique (as usual, because the difference of the $F$'s is a decreasing function and the $c$ is increasing). We show that $\eta^*(S) > \lambda_n^*(S)$. In fact, if $\eta^*(S) \le \lambda_n^*(S)$, then $\lambda_j^*(S,\eta^*(S))=\lambda_j^*(S)$ for each $j\in N$, and we have shown in Step 1, that the equality in \eqref{eq9} does not old for $\lambdab^*(S)$.

\textbf{Step 3:} We just need to show that this is a Nash equilibrium of $\Gamma(S,\eta^*(S))$. At first, observe that no agent has incentives to deviate to a quantity in $(\eta^*(S), 1/\sigma^2]$, because of the convexity of $\Phi$. It remains to be shown that no agent has incentives to deviate to zero. Agent $s$ does not have incentives by \eqref{eq9}. For any other agent $i\neq n$, s.t. $\lambda_i^*(S,\eta^*(S)) = \lambda_s^*(S,\eta^*(S))$, if agent $s$, who is the most privacy concerned, does not have incentives to deviate from $\lambda_i^*(S,\eta^*(S))$, that is still valid for $i$. For any other agent $i\neq n$, s.t. $\lambda_i^*(S,\eta^*(S)) > \lambda_s^*(S,\eta^*(S))$, if $i$ does not have incentives to deviate to $\eta^*(S)$, then, because of the convexity of the cost function, she cannot have incentives to deviate to $0$.

\subsection{Proof of Theorem~\ref{opteta1stageHet}}\label{opteta1stageHetproof}
At first, because of Proposition~\ref{DecEstCostHet}, for each $S\subseteq N$, $\sigma_M^2(\lambdab^*(S)) \ge \sigma_M^2(\lambdab^*(N))$. Then, for the analyst it is more convenient to have the complete set of agents playing. Moreover, from the KKT conditions in Equation \eqref{KKT2}, when implementing $\Gamma$
\begin{align*}
c_n'(\lambda_n^*(N)) = \frac{F'(\sigma_M^2(\lambdab^*(N)))}{(\sum_{j\in N}\lambda_j^*(N,))^2},
\end{align*}
as we assumed that $\lambda_n^*(N) \neq 1/\sigma^2$ (otherwise the estimation would have been already optimal). When implementing $\Gamma(N,\eta^*(N))$, or we have that $\lambda_n^*(N,\lambda^*(N,\eta^*(N)))=1/\sigma^2$, and in this case we have proved our result. In fact, it follows that every agent is playing $1/\sigma^2$ and that the estimation is now optimal. If  $\lambda_n^*(N,\lambda^*(N,\eta^*(N))) \neq 1/\sigma^2$, then
\begin{align}\label{eq11}
c_n'(\lambda_n^*(N,\lambda^*(N,\eta^*(N)))) = \frac{F'(\sigma_M^2(\lambdab^*(N,\eta^*(N))))}{(\sum_{j\in N}\lambda_j^*(N,\eta^*(N)))^2}.
\end{align}
As
$$
\lambda_n^*(N,\lambda^*(N,\eta^*(N))) > \eta^*(N) > \lambda_n^*(N),
$$
it follows that
\begin{align}\label{eq12}
c_n'(\lambda_n^*(N,\lambda^*(N,\eta^*(N)))) > c_n'(\lambda_n^*(N)),
\end{align}
because of the convexity of $c_n$. Assume by contradiction that
$$
\sigma_M^2(\lambdab^*(N,\eta)) \ge \sigma_M^2(\lambdab^*(N)),
$$
it follows that
$$
\frac{F'(\sigma_M^2(\lambdab^*(N,\eta^*(N))))}{(\sum_{j\in N}\lambda_j^*(N,\eta^*(N)))^2} \ge \frac{F'(\sigma_M^2(\lambdab^*(N)))}{(\sum_{j\in N}\lambda_j^*(N,))^2},
$$
and this contradicts \eqref{eq12}.

\subsection{Proof of Lemma~\ref{theorem2stage2nd}}\label{theorem2stage2ndproof}

$\Gamma^P(\eta)$ is still a potential game, with potential function $\Phi$ as in \eqref{potentialfunction1stage}, but defined on the domain $[\eta,1/\sigma^2]^p$. The set of Nash equilibria of $\Gamma^P(\eta)$ is included in the set of the local minima of $\Phi$ on this new domain. The unique local minimum of $\Phi$ is given by $\lambdab^*(\pb,\eta) = \lambdab^*(p)$, if $\lambda^*(p)\le \eta$, and by $\lambdab^*(\pb,\eta) =\etab$ otherwise. In both the cases, this is a Nash equilibrium, because of the convexity of the potential function (any deviation of an agent would make her cost function bigger).

\subsection{Proof of Lemma~\ref{theorem2stage1st}}\label{theorem2stage1stproof}
Because of Lemma~\ref{theorem2stage2nd}, given a vector $\pb$ in the first stage, in the second stage the agents in $P$ are going to choose a precision level as in \eqref{solution2ndstage}.

At first, we observe that $(1,\ldots ,1)$ is a Nash equilibrium when $\eta\in [\lambda^*(n-1),\eta^*(n)]$. As $\lambda^*(n) < \lambda^*(n-1) \le \eta$, if $\pb=(1,\ldots ,1)$, in the second stage, the agents are going to play $\eta$ at equilibrium, and if an agent decides to deviate to $p_i=0$, the remaining $n-1$ agents are still going to play $\eta$ at equilibrium. Then, by deviating to $p_i=0$, agent $i$ cannot make her cost function smaller, as
$$
\frac{1}{n\eta}+c\eta^k\le \frac{1}{(n-1)\eta}
$$
by \eqref{fixedpoint3eta} as $\eta\le \eta^*(n)$, where the left term represents her cost before deviation, and the right one represents her cost after deviation.

To prove that there are no other Nash equilibria, let $\eta \in [\lambda^*(n-1),\eta^*(n)]$. Suppose by contradiction that there exists an equilibrium s.t. the set $N\setminus P$ of agents who choose zero in the first stage is non-empty. Then, the agents in $P$ choose $\lambda^*(p,\eta)$ at equilibrium in the second stage. Then, if $\lambda^*(p)< \lambda^*(p-1) < \eta$, then an agent in $N\setminus P$ has incentives to deviate choosing $p_i=1$. The same happens if $\eta < \lambda^*(p)< \lambda^*(p-1)$. While if $\lambda^*(p)<\eta <\lambda^*(p-1)$, then the agents in $P$ have incentives to deviate choosing $p_i=0$. 
%Suppose that player $i\in N$ has an incentive to deviate to $p_i=0$, this means that
%\begin{align}\label{eqproof1}
%& \sigma_M^2(\lambdab^*(n-1),\eta)<c(\lambda^*(n,\eta))^k+\sigma_M^2(\lambdab^*(n,\eta)).
%\end{align}
%Now, we observe that when $\eta \in[0,\lambda^*(n)]$, because of \eqref{solution2ndstage} we have that $\lambda^*(n,\eta)=\lambda^*(n)$ and as $\lambda^*(n-1,\eta)\ge \lambda^*(n,\eta)$, $\lambda^*(n-1,\eta)= \lambda^*(n-1)$. Then \eqref{eqproof1} is equivalent to
%\begin{align*}
%& \sigma_M^2(\lambdab^*(n-1))<c(\lambda^*(n)^k+\sigma_M^2(\lambdab^*(n)),
%\end{align*}
%where the left term is the cost function $J_i$ evaluated when each $j\in N\setminus \{i\}$ plays $\lambda^*(n)$ and $i$ plays $0$, while the right term is $J_i$ at equilibrium $\lambdab^*(n)$. And this contradicts the fact that $\lambdab^*(n)$ is an equilibrium for the game $\Gamma$.
%
%\textit{to be finished...}

\subsection{Proof of Theorem~\ref{opteta2stage}}\label{opteta2stageproof}

At first, we observe that, because of Lemma \ref{theorem2stage1st}, $\sigma_M^2(\lambda^*(n,\eta^*(n))) < \sigma_M^2(\lambda^*(n,\eta))$ for each $\eta\in [\lambda^*(n-1),\eta^*(n))$. In fact, when $\eta\in [\lambda^*(n-1),\eta^*(n)]$, at the unique equilibrium, every agent is choosing to participate in the first stage and then she is choosing the same non-zero precision level $\lambda^*(n,\eta)$ and then the estimation has minimum cost when this precision level is maximal, i.e. when it is equal to $\eta^*(n)$.

When $\eta\in [0,\lambda^*(n)]$, then for every vector $\pb\in \{0,1\}^n$ in the first stage, in the second stage the agents in $N\setminus P$ choose a precision level $\lambda^*(n-p)$ and estimation cost is $\sigma_M^2(\lambda^*(n-p))\ge \sigma_M^2(\lambda^*(n))$ because of Corollary \ref{DecEstCost}. When $\eta\in [\lambda^*(n),\lambda^*(n-1)]$, then for every vector $\pb\in \{0,1\}^n$ with $p\le n-1$ we have again, as before, a non-optimal estimation, while we show that $\pb=(1,\ldots ,1)$ is not a Nash equilibrium. In fact,
\begin{align*}
& \frac{1}{(n-1)\lambda^*(n-1)} < \frac{1}{n \eta}+c \eta^k
\end{align*}
for each $k\ge 2$, and this means that each agent can make her cost function smaller by deviating to zero.

Finally, when $\eta\in [\eta^*(n),1/\sigma^2]$, for every vector $\pb\in \{0,1\}^n$ in the first stage, in the second stage the agents in $N\setminus P$ choose a precision level equal to $\lambda^*(n-p)$ or equal to $\eta$ and, as we have already seen in the proof of Theorem \ref{opteta1stage}, this does not provide a minimum value for the estimation cost.

\subsection{Proof of Theorem~\ref{costtheorem}}\label{costtheoremproof}

We prove at first that $J_A(n)$ is a definitely increasing sequence, i.e., that $J_A(n) > J_A(n-1)$, implies $J_A(n+1) > J_A(n)$. We have that
\begin{align*}
& J_A(n) > J_A(n-1) \\
\Leftrightarrow & F\left(\frac{1}{n\eta^*(n)}\right) + Cn > F\left(\frac{1}{(n-1)\eta^*(n-1)}\right) + C(n-1) \\
\Leftrightarrow & C > F\left(\frac{1}{(n-1)\eta^*(n-1)}\right) - F\left(\frac{1}{n\eta^*(n)}\right).
\end{align*}
As the right term decreases when $n$ increases, and as the left term is a constant, it follows that this inequality is definitely true while $n$ increases. Looking for the optimal $n^*$, we need to find the highest $n$ s.t. the previous inequality does not hold, i.e., s.t.
$$
C \le F\left(\frac{1}{(n-1)\eta^*(n-1)}\right) - F\left(\frac{1}{n\eta^*(n)}\right).
$$
It is now sufficient to observe that the term on the right is equal, by definition of $\eta^*(n)$ to $c(\eta^*(n))$. The highest $n$ for which this inequality holds, is the optimal number of agents $n^*$. If this inequality is never satisfied, it means that the estimation cost is increasing in $n$, and than the optimum number of agents is $1$.

\end{document}